\documentclass[twocolumn]{aastex631}

\def\msun{M$_{\odot}$}
\def\arc{^{\prime\prime}}
\def\zphot{\ifmmode z_{\rm phot}\else$z_{\rm phot}$\fi}

\def\Hb{H$\beta$}

\def\Oiiilong{[O\,{\sc iii}]\,$5008\,{\rm \AA}$}
\def\Oiiishort{[O\,{\sc iii}]\,$4959\,{\rm \AA}$}
\def\Oiiiboth{[O\,{\sc iii}]\,$\lambda\lambda4959,\,5008\,{\rm \AA}$}
\def\Hb{H$\beta$}

\def\ltsima{$\buildrel<\over\sim$}
\def\la{\lower.5ex\hbox{\ltsima}~}
\def\gtsima{$\buildrel>\over\sim$}
\def\ga{\lower.5ex\hbox{\gtsima}~}
\def\deg~{$^{\circ}$}

\begin{document}

\title{SAPPHIRES: A Galaxy Over-Density in the Heart of Cosmic Reionization at $z=8.47$}

\correspondingauthor{Yoshinobu Fudamoto}
\email{yoshinobu.fudamoto@gmail.com, y.fudamoto@chiba-u.jp}

\author[0000-0001-7440-8832]{Yoshinobu Fudamoto} 
\affiliation{Center for Frontier Science, Chiba University, 1-33 Yayoi-cho, Inage-ku, Chiba 263-8522, Japan}
\affiliation{Steward Observatory, University of Arizona, 933 N Cherry Avenue, Tucson, AZ 85721, USA}

\author[0000-0003-4337-6211]{Jakob M. Helton}
\affiliation{Steward Observatory, University of Arizona, 933 N Cherry Avenue, Tucson, AZ 85721, USA}

\author[0000-0001-6052-4234]{Xiaojing Lin}
\affiliation{Department of Astronomy, Tsinghua University, Beijing 100084, China}
\affiliation{Steward Observatory, University of Arizona, 933 N Cherry Avenue, Tucson, AZ 85721, USA}

\author[0000-0002-4622-6617]{Fengwu Sun}
\affiliation{Center for Astrophysics $|$ Harvard \& Smithsonian, 60 Garden St., Cambridge, MA 02138, USA}

\author[0000-0002-2517-6446]{Peter Behroozi}
\affiliation{Department of Astronomy and Steward Observatory, University of Arizona, Tucson, AZ 85721, USA}

\author[0000-0003-4512-8705]{Tiger Yu-Yang Hsiao}
\affiliation{Center for Astrophysics $|$ Harvard \& Smithsonian, 60 Garden St., Cambridge, MA 02138, USA}
\affiliation{Center for Astrophysical Sciences, Department of Physics and Astronomy, The Johns Hopkins University, 3400 N Charles St. Baltimore, MD 21218, USA}
\affiliation{Space Telescope Science Institute (STScI), 3700 San Martin Drive, Baltimore, MD 21218, USA}

\author[0000-0003-1344-9475]{Eiichi Egami}
\affiliation{Steward Observatory, University of Arizona, 933 N Cherry Avenue, Tucson, AZ 85721, USA}

\author[0000-0002-8651-9879]{Andrew J.\ Bunker}
\affiliation{Department of Physics, University of Oxford, Denys Wilkinson Building, Keble Road, Oxford OX1 3RH, U.K.}

\author[0000-0003-2388-8172]{Francesco D'Eugenio}
\affiliation{Kavli Institute for Cosmology, University of Cambridge, Madingley Road, Cambridge CB3 0HA, UK}
\affiliation{Cavendish Laboratory, University of Cambridge, 19 JJ Thomson
Avenue, Cambridge CB3 0HE, UK}

\author[0000-0002-6047-430X]{Yuichi Harikane}
\affiliation{Institute for Cosmic Ray Research, The University of Tokyo, 5-1-5 Kashiwanoha, Kashiwa, Chiba 277-8582, Japan}

\author[0000-0002-1049-6658]{Masami Ouchi}
\affiliation{National Astronomical Observatory of Japan, 2-21-1 Osawa, Mitaka, Tokyo 181-8588, Japan}
\affiliation{Institute for Cosmic Ray Research, The University of Tokyo, 5-1-5 Kashiwanoha, Kashiwa, Chiba 277-8582, Japan}
\affiliation{Department of Astronomical Science, SOKENDAI (The Graduate University for Advanced Studies), Osawa 2-21-1, Mitaka, Tokyo, 181-8588, Japan}
\affiliation{Kavli Institute for the Physics and Mathematics of the Universe (WPI), University of Tokyo, Kashiwa, Chiba 277-8583, Japan}

\author[0000-0003-4247-0169]{Yichen Liu} 
\affiliation{Steward Observatory, University of Arizona, 933 N Cherry Avenue, Tucson, AZ 85721, USA}

\author[0000-0003-3762-7344]{Weizhe Liu}
\affiliation{Steward Observatory, University of Arizona, 933 N Cherry Avenue, Tucson, AZ 85721, USA}

\author[0000-0002-4985-3819]{Roberto Maiolino}
\affiliation{Kavli Institute for Cosmology, University of Cambridge, Madingley Road, Cambridge CB3 0HA, UK}
\affiliation{Cavendish Laboratory, University of Cambridge, 19 JJ Thomson
Avenue, Cambridge CB3 0HE, UK}
\affiliation{Department of Physics and Astronomy, University College London,
Gower Street, London WC1E 6BT, UK}

\author[0000-0001-7673-2257]{Zhiyuan Ji}
\affiliation{Steward Observatory, University of Arizona, 933 N Cherry Avenue, Tucson, AZ 85721, USA}
 
\author[0000-0002-5768-738X]{Xiangyu Jin}
\affiliation{Steward Observatory, University of Arizona, 933 N Cherry Avenue, Tucson, AZ 85721, USA}

\author[0000-0003-0747-1780]{Wei Leong Tee}
\affiliation{Steward Observatory, University of Arizona, 933 N Cherry Avenue, Tucson, AZ 85721, USA}

\author[0000-0002-7633-431X]{Feige Wang}
\affiliation{Department of Astronomy, University of Michigan, 1085 S. University Ave., Ann Arbor, MI 48109, USA}

\author[0000-0001-9262-9997]{Christopher N. A. Willmer}
\affiliation{Steward Observatory, University of Arizona, 933 N Cherry Avenue, Tucson, AZ 85721, USA}

\author[0000-0002-5768-8235]{Yi Xu}
\affiliation{Institute for Cosmic Ray Research, The University of Tokyo, 5-1-5 Kashiwanoha, Kashiwa, Chiba 277-8582, Japan}
\affiliation{Department of Astronomy, Graduate School of Science, the University of Tokyo, 7-3-1 Hongo, Bunkyo, Tokyo 113-0033, Japan}

\author[0000-0003-3307-7525]{Yongda Zhu}
\affiliation{Steward Observatory, University of Arizona, 933 N Cherry Avenue, Tucson, AZ 85721, USA}




\begin{abstract}

We report the discovery of a galaxy proto-cluster candidate (dubbed MACS0416-OD-z8p5) at a spectroscopic redshift of $z\sim8.47$, dating back to $\sim550\,{\rm Myr}$ after the Big Bang. The observations are part of the JWST Cycle-3 treasury program, Slitless Areal Pure-Parallel HIgh-Redshift Emission Survey (SAPPHIRES) with NIRCam-grism. Using wide field slitless spectroscopy (WFSS) obtained in the MACS0416 parallel field, we robustly confirm nine galaxies at $z_{\rm spec}\sim8.47$ via emission line detections of \Oiiilong\ (with $>5\,\sigma$) and tentatively confirm one additional galaxy (at $\sim3\,\sigma$).
This discovery represents the highest-redshift, spectroscopically confirmed galaxy over-density known to date, which is $\sim6$--$8$ times more dense than the average volume density of galaxies at the same redshift.
Furthermore, a galaxy hosting a low-mass active galactic nucleus (``Little-Red-Dot'') is found as a member, suggesting an early emergence of active, massive black holes and feedback between these black holes and their surrounding environments.
We also discuss the spatial structures connecting the galaxy over-density to nearby massive star-forming galaxies (separated by $\sim 5\,{\rm pM pc}$), including MACS0416-Y1 and MACS0416-JD.
This finding of a massive dark matter halo hosting a galaxy over-density at $z\sim8.5$ is surprising given that our survey covered only a small, random field ($16.5\,{\rm arcmin^2}$) as part of a pure parallel observation. 
The comparison with cosmological simulations shows that the likelihood of finding such a large-scale structure is $<5\,\%$ under the current galaxy formation scenario and the observed survey volume.
Our results demonstrate the power of WFSS observations to build a complete line-emitter sample and suggest an important role for over-densities in enhancing galaxy formation by funneling large-scale gas supplies into small cosmological volumes.
\end{abstract}



\section{Introduction} \label{sec:intro}
After the first galaxies and stars formed, the ultraviolet (UV) light they emitted began ionizing pervasive neutral hydrogen in the intergalactic space, known as cosmic reionization \citep[for reviews, see][]{Brom2011,Madau2014}.
Understanding galaxy formation at redshifts of $z\gtrsim6$ and its role in the cosmic reionization is one of the key goals of current extragalactic astronomy \citep[e.g.,][]{Planck2016,Wise2019,Robertson2022}. 
Over the past decades, it has become evident that cosmic reionization started at $z\sim20$ -- $15$, reached its midpoint at $z\sim7.7$, and was completed by redshift $z\sim5-6$ \citep[e.g.,][]{Fan2006, Hoag2019}.
Nevertheless, the exact onset of reionization and the detailed role of early galaxy evolution remain areas of active investigation.

Thanks to the unprecedented capability of the James Webb Space Telescope (JWST), recent studies identified large numbers of star-forming galaxies out to redshifts of $z>14$ \citep[][]{Castellano2022,Finkelstein2022,Naidu2022,Bunker2023,Carniani2024,Curtis-Lake2023,Donnan2023,Harikane2023,Carniani2024b,Roberts-Borsani2024,Hainline2024,Hsiao2023,Hsiao2024,Witstok2024a,Schouws2024,Wang2024}. Although they are found to be very luminous and numerous, individual star-forming galaxies alone may be inefficient to fully ionize the surrounding large-scale neutral intergalactic medium \citep[IGM; e.g.,][]{Dayal2009,Larson2022}.
On the other hand, dense regions of galaxies, or proto-clusters, are alternatively thought to be crucial to start the reionization of the IGM and to create large ionized bubbles extending $\sim$ a few comoving Mpc \citep[e.g.,][]{McQuinn2016}. The ionized bubbles created by multiple galaxies further allow ionizing photons to travel larger distances and lead to reionization over larger scale in the Universe \citep[e.g.,][]{Tilvi2020,Endsley2021, Leonova2022,Jin2024,Saxena2023,Witstok2024b,Witstok2024a,Whitler2024,Witstok2025}.

At the same time, dense galaxy environments are crucial for understanding galaxy evolution. In galaxy proto-clusters, enhanced gas accretion and galaxy-galaxy interactions may drive galaxy growth, resulting in the cosmic star-formation activity being dominated by over-dense environments before Cosmic Noon \citep{Chiang2017,Lim2024}.
Such accelerated galaxy evolution within over-dense environments is observationally supported by several spectroscopic and photometric studies \citep{Toshikawa2014,Higuchi2019,Harikane2019,Jin2023,Hashimoto2023,Champagne2024a,Helton2024,Morishita2024}.
Some of these studies even show that, if the galaxy over-densities are too massive, galaxy proto-clusters may be inefficient at producing escaping ionizing photons due to their largely accumulated neutral gas and enhanced dust production, both of which are highly responsible for preventing ionizing photons from escaping from galaxies \citep[][]{Hashimoto2023,Ma2024}.
Thus, constraining and understanding galaxy growth in galaxy proto-clusters are essential to determine both the progression of galaxy evolution as well as the history of the Universe.



JWST/NIRCam's wide-field slitless spectroscopy (WFSS) is proven to be extremely efficient to identify and study complete galaxy samples \citep[e.g.,][]{Oesch2023,Sun2023}.
Studies using WFSS have now confirmed several galaxy over-densities at $z\sim5$ -- $8$ \citep[e.g.,][]{Champagne2024a,Helton2024,Helton2024b,Ma2024,Sun2024,Eilers2024,Herard-Demanche2025}, enabling unbiased investigations of environmental effects on the galaxy evolution process at high redshifts.
However, samples of very high-redshift galaxy over-densities are still limited. Thus, it is imperative to further exploit galaxy over-densities at high redshift and investigate their nature.

In this study, we report the discovery and in-depth analysis of a galaxy over-density at $z=8.47$ (MACS0416-OD-z8p5), the highest redshift record for a spectroscopically confirmed galaxy over-density or proto-cluster candidate known to date.
We present detailed properties of member galaxies to study their evolutionary stage using spectroscopic and photometric data.

This paper is organized as follows: in \S\ref{sec:observation} we describe our observations and data used in this study. In \S\ref{sec:analysis}, we present our data analysis and measurements.  In \S\ref{sec:results}, we present the results of the study. In \S\ref{sec:discussion}, we discuss our results. Finally, we conclude with the summary in \S\ref{sec:conclusion}.
Throughout this paper, we assume a cosmology with $(\Omega_m,\Omega_{\Lambda},h)=(0.3,0.7,0.7)$.
All magnitudes are in the AB system \citep{1983ApJ...266..713O}.
With the assumed cosmology, $1\arc$ at $z=8.47$ corresponds to $4.6\,{\rm pkpc}$ .

\begin{figure*}
    \centering
    \includegraphics[width=0.9\textwidth]{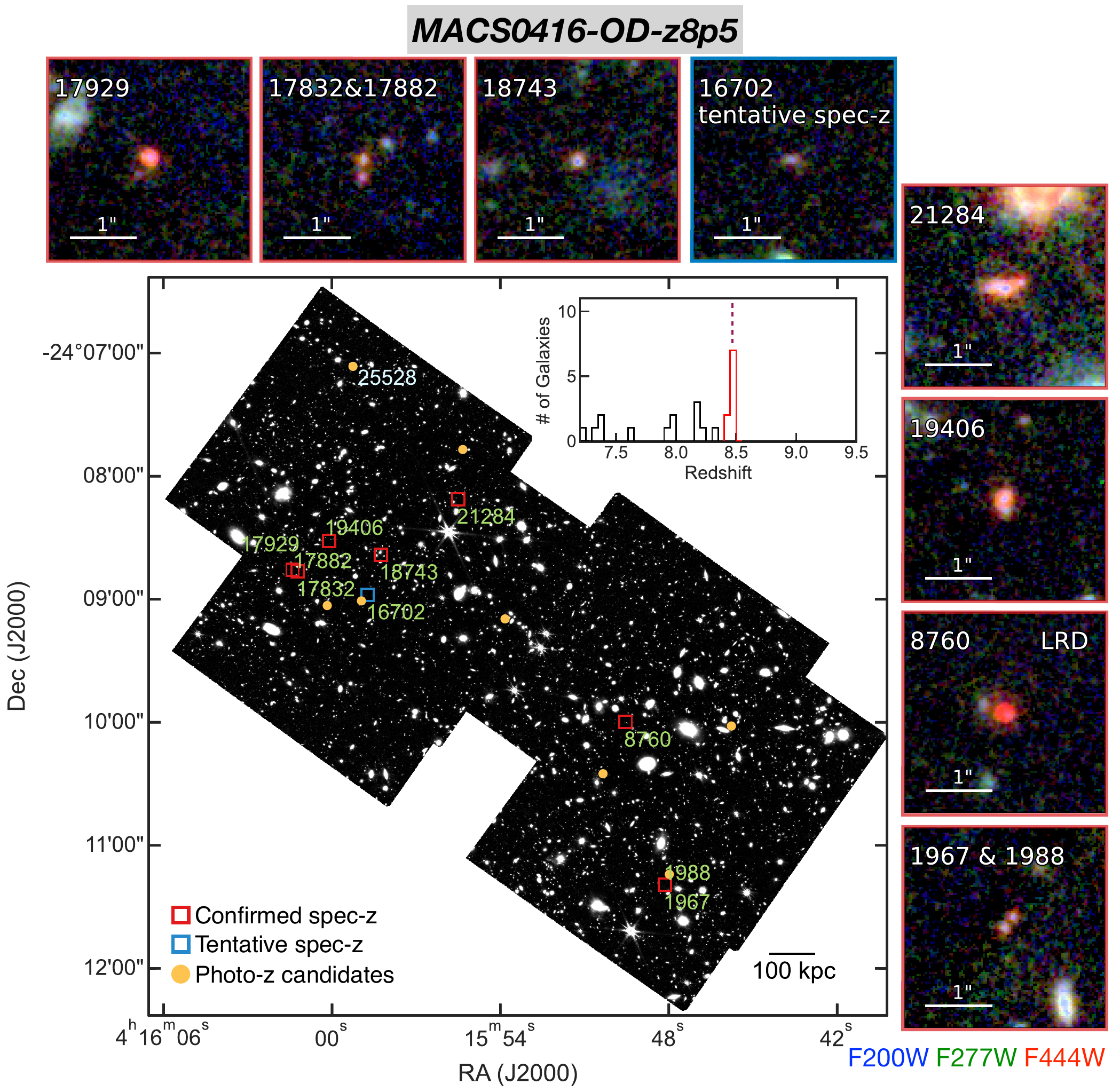}
    \caption{Spatial distribution of MACS0416-OD-z8p5 and false-color images of its member galaxies.
    The large central panel shows the sky distribution of MACS0416-OD-z8p5 overlaid on the SAPPHIRES-EDR F444W mosaic (north is up, east is left). Red squares indicate the positions of spectroscopically confirmed member galaxies. The inset histogram shows number counts of spectroscopically confirmed galaxies in the SAPPHIRES-EDR.
    Surrounding smaller panels show $3^{\prime\prime}\times3^{\prime\prime}$ false-color images of each member galaxy, created using F200W, F277W, and F444W filters. A blue square shows a member galaxy candidate  with a tentative spectroscopic redshift, while yellow points represent member galaxy candidates selected based on their secure photometric redshifts.
    The distribution of MACS0416-OD-z8p5 member galaxies across the F444W image may suggest that the over‑density extends beyond the current survey footprint.
    }
    \label{fig:entire}
\end{figure*}

\section{Observations} \label{sec:observation}
This study is based on the JWST Cycle‑3 Treasury Survey,  Slitless Areal Pure‑Parallel HIgh‑Redshift Emission Survey (SAPPHIRES; GO‑6434; PI: E. Egami). SAPPHIRES performs imaging and WFSS using $711\,{\rm hours}$ of pure‑parallel NIRCam observations. Within the SAPPHIRES data, we focus on the SAPPHIRES Early Data Release (EDR), which was obtained in a parallel field of the MACS0416 lensing cluster. Observations, photometry, and spectroscopy for the SAPPHIRES‑EDR are presented in \citet{Sun2025}; interested readers are referred to the EDR paper for details.

In summary, with a total dual-channel exposure time of $47.2\,{\rm hours}$, the SAPPHIRES‑EDR obtained NIRCam imaging in 13 medium‑ and broad‑band filters, as well as F356W and F444W NIRCam WFSS data in both the orthogonal Grism‑C and Grism‑R dispersion directions. Typically, the imaging observations reached $5\,\sigma$ sensitivities of $\sim29\,{\rm AB\ mag}$ for broad-band filters.
The grism observations achieved a typical $5\,\sigma$ emission‑line sensitivity of $9\times10^{-19}\,{\rm erg\,s^{-1}\,cm^{-2}}$ at $4.7\,{\rm \mu m}$ (i.e., the expected wavelength of \Oiiilong\ at $z\sim8.5$) for the NIRCam module A and is $\sim20\,\%$ less sensitive for module B. The emission line sensitivity typically corresponds to a star formation rate (SFR) sensitivity of $\sim1\,{\rm M_{\odot}\,yr^{-1}}$ based on previous \Oiiilong\ observations \citep{Villa-Velez2021,Meyer2024}.

Although the observed field is next to the lensing cluster MACS0416, the field is substantially distant ($\gtrsim4^{\prime}$) from the lensing cluster.
Based on the cluster mass model of \citet{richard_2021}, we conclude that the gravitational magnification effects from the cluster are negligible ($\mu\lesssim1.1$).
 
\section{Analysis} \label{sec:analysis}

\subsection{Spectroscopic Galaxy Identifications}
The galaxy over-density (dubbed MACS0416-OD-z8p5, hereafter) was identified from the spectroscopic redshift samples of the SAPPHIRES-EDR.
We constructed the spectroscopic sample by combining photometric redshifts with emission‑line detections from the grism spectra.
In particular, we estimated photometric redshifts using the template‑fitting code \texttt{eazy} \citep{Brammer2008} with high‑redshift galaxy templates configured as in \citet{Hainline2024}.
To confirm the spectroscopic redshifts, we searched for combinations of significant ($>4$--$5\sigma$) emission lines in the grism spectra using both the grism-R and grism-C dispersion directions simultaneously.
All spectroscopic redshifts and spectra were visually inspected in the collaboration (see details in \citealt{Sun2025}). 

Within the final spectroscopic sample, there are $16$ galaxies robustly selected at $z>8$, $10$ of which are located at $z=8.427$ -- $8.479$. A clear concentration of galaxies at $z\sim8.4$ is evident from these results, enabling us to identify the galaxy over-density MACS0416-OD-z8p5 at $z_{\rm median}=8.473$.
Figure \ref{fig:entire} shows their spatial distribution and false color images. Nine spectroscopic member galaxies are mostly distributed around the north-east part of the footprint, while a few galaxies are located extending towards the south-west corner. Furthermore, the figure includes photometric candidates (see \S\ref{sec:photoz}) and a candidate with an active galactic nuclei (see \S\ref{sec:AGN}). See also \S\ref{sec:extent} for detailed discussions of spatial structure and extent.
Table \ref{tab:galaxies} summarizes the basic properties of the spectroscopic member galaxies as well as a tentative spectroscopic sample that will be presented in \S\ref{sec:photoz}.

Among the nine galaxies at $z\sim8.4$, we note some are located very close to each other, with separations of only $\sim 0\farcs2$ or $0.9\,{\rm pkpc}$ at $z=8.47$ (ID-17832/ID-17882 and ID-1963/ID-1988, see Fig. \ref{fig:entire}).
Rather than counting these close pairs as single galaxies, one could interpret them as sub-components of single galaxy structures.
Nevertheless, reducing the number of member galaxies from 9 to 7 does not change our conclusion (e.g., our interpretations about over-density properties).

\begin{table*}[t]
    \centering
    \renewcommand{\tabcolsep}{0.12cm}
    \caption{Spectroscopically confirmed galaxy members of the SAPPHIRES-z8-OD0 and their measured properties.}
    \begin{tabular}{lccccccccc}
    \hline
       ID  &  RA & Dec &  $z_{\rm spec}$ & ${\rm mag_{F200W}}$ &  ${\rm mag_{F444W}}$ & ${{\rm FWHM}_{H\beta}}$ & ${{\rm FWHM}_{\rm [OIII]5007}}$ & ${10^{-19} \,f_{H\beta}}$ & ${10^{-19}\,f_{\rm [OII]5007}}$\\
       & deg & deg & & ABmag. & ABmag. & ${\rm km/s}$ & ${\rm km/s}$ & ${\rm erg/s/cm^2}$ & ${\rm erg/s/cm^2}$ \\
    \hline
        1967$^{a}$ & 63.9515233 & -24.1895010 & 8.471 & $29.84\pm0.46$& $27.97\pm0.03 $  & -- & $270 \pm 130$ & $<25$ & $21.1\pm5.2$ \\
    1988$^{a}$ & 63.9514815 & -24.1894572 & 8.471 & $29.82\pm0.41$& $28.25\pm0.04 $  & 
    -- & $540\pm110$ & $<44$ & $27.8\pm4.1$ \\
    8760 & 63.9573331 & -24.1674685 & 8.479 & $28.04\pm0.08$& $26.14\pm0.01 $  & $1010 \pm 190$ & $110\pm70$ & $32.4\pm5.0$ & $17.7\pm2.5$ \\
    17832$^{b}$ & 64.0060246 & -24.1471058 & 8.473 & $29.65\pm0.42$& $28.22\pm0.04 $  & -- & $150\pm80$ & $<5.9$ & $12.0\pm2.2$ \\
    17882$^{b}$ & 64.0060195 & -24.1470406 & 8.473 & $29.96\pm0.60$& $28.06\pm0.04 $ & -- & $300\pm70$ & $<12$ & $19.2\pm2.5$ \\
    17929 & 64.0067601 & -24.1468564 & 8.473 & $27.81\pm0.07$& $26.70\pm0.01 $  & $100\pm220$ & $146\pm20$ & $8.5\pm1.6$ & $50.7\pm2.4$ \\
    18743 & 63.9936700 & -24.1448186 & 8.483 & $28.18\pm0.08$& $27.85\pm0.03 $  & -- & $180\pm70 $ & $<6.7$ & $15.8\pm2.3$ \\
    19406 & 63.9609227 & -24.1431647 & 8.427 & $27.80\pm0.13$& $27.66\pm0.07 $  & $120\pm80$ & $170\pm20$ & $13.2\pm2.4$ & $61.7\pm2.8$ \\
    21284$^{c}$ & 63.9821549 & -24.1372869 & 8.433 & $26.64\pm0.04$& $25.99\pm0.01 $  & -- & -- & $<13$ & $65.3\pm4.0$ \\
    \hline
    \multicolumn{10}{c}{Tentative Spectroscopic Sample}\\
    \hline
    16702 & 63.9956191 & -24.1502346 & 8.47 & $28.90\pm0.15$ & $28.36\pm0.05$ & -- & -- & $<7.3$ & $6.4\pm1.6$ \\
    \hline
    \multicolumn{6}{l}{$^{a}$ $^{b}$ Close pair of galaxies.}\\
    \multicolumn{6}{l}{ $^{c}$ Lines are contaminated by nearby sources.}\\
    \end{tabular}
    \label{tab:galaxies}
\end{table*}

\subsection{Photometric Galaxy Searches}
\label{sec:photoz}
In addition to the spectroscopic sample, we searched for candidate member galaxies using their photometric redshifts and tentative emission lines detected in their spectra.
The photometric redshift-based searches aim to identify galaxies that, despite being faint  in emission lines, show features placing them at $z\sim8$ -- $9$, such as a Ly$\alpha$ break between the F090W and F150W images and flux excess in the F444W images resulting from weak contamination by emission lines.
We expect that these line-faint galaxies include temporarily quenched galaxies \citep[e.g.,][]{Looser2024,Witten2025}, dust-obscured galaxies \citep[e.g.,][]{Fudamoto2021,Barrufet2023}, and low-mass galaxies that are below our robust line detection limits.

Similar to \citet{Helton2024}, we selected photometric candidates using their accurate photometric redshifts according to the following four criteria: ({\sc i}) both $z_{\rm map}$ (the maximum likelihood redshift) and $z_{\rm med}$ (the median redshift obtained using \texttt{eazy}) lie within $8.47\pm0.2$. ({\sc ii}) The F444W magnitude is brighter than $29.1\,{\rm mag}$, which corresponds to the median $5\,\sigma$ limiting magnitude of the F444W image. ({\sc iii}) The $16\,{\rm th}$ -- $84\,{\rm th}$ percentile interval of photometric redshift is smaller than $20\%$ of $1 + z_{\rm med}$.
({\sc iv}) The $5\,{\rm th}$ -- $95\,{\rm th}$ percentile interval is smaller than $40\%$ of $1 + z_{\rm med}$.

Using these criteria, we selected $17$ galaxies. Among these, $4$ galaxies are already spectroscopically confirmed as members in this work, while $5$ additional galaxies have spectroscopic redshifts ranging $z_{\rm spec}=6.29$ to $8.15$, indicating that they are not members \citep{Sun2025}.
Among the remaining photometric sample, one galaxy (ID-16702) shows a weak $\sim 3\,\sigma$ signal at the expected wavelength of \Oiiilong\ from $z=8.482$ (Figure \ref{fig:spectra_summary}). Indeed, ID-16702 has a photometric redshift consistent with $z_{\rm map}=8.48$, suggesting it is highly likely to be a member of MACS0416-OD-z8p5.
Finally, we identified $7$ photometric candidates that lack any credible emission line detections in their spectra (see Table \ref{tab:photoz} in the Appendix). 

In our following analysis, we treat only ID-16702 as a tentative spectroscopic candidate and consider it as a member of MACS0416-OD-z8p5. However, the other $7$ photometric candidates are regarded solely as potential candidates and are not formally counted as members of the over-density when analyzing its properties.
Although our photometric redshifts are relatively accurate, spectroscopic redshifts remain essential for confirmation, especially since $5$ galaxies have been spectroscopically confirmed to lie at different redshifts.

Compared to the final spectroscopic sample, the photometric candidates generally have higher dust attenuation and older ages (see Table \ref{tab:photoz}). These features suggest that we might be missing such highly dust-attenuated and/or older stellar population dominated galaxies, that are faint in emission lines.
Nevertheless, future spectroscopic confirmations are essential to investigate this further.

\subsubsection*{ID-25528}

\begin{figure}
    \centering
    \includegraphics[width=0.99\linewidth]{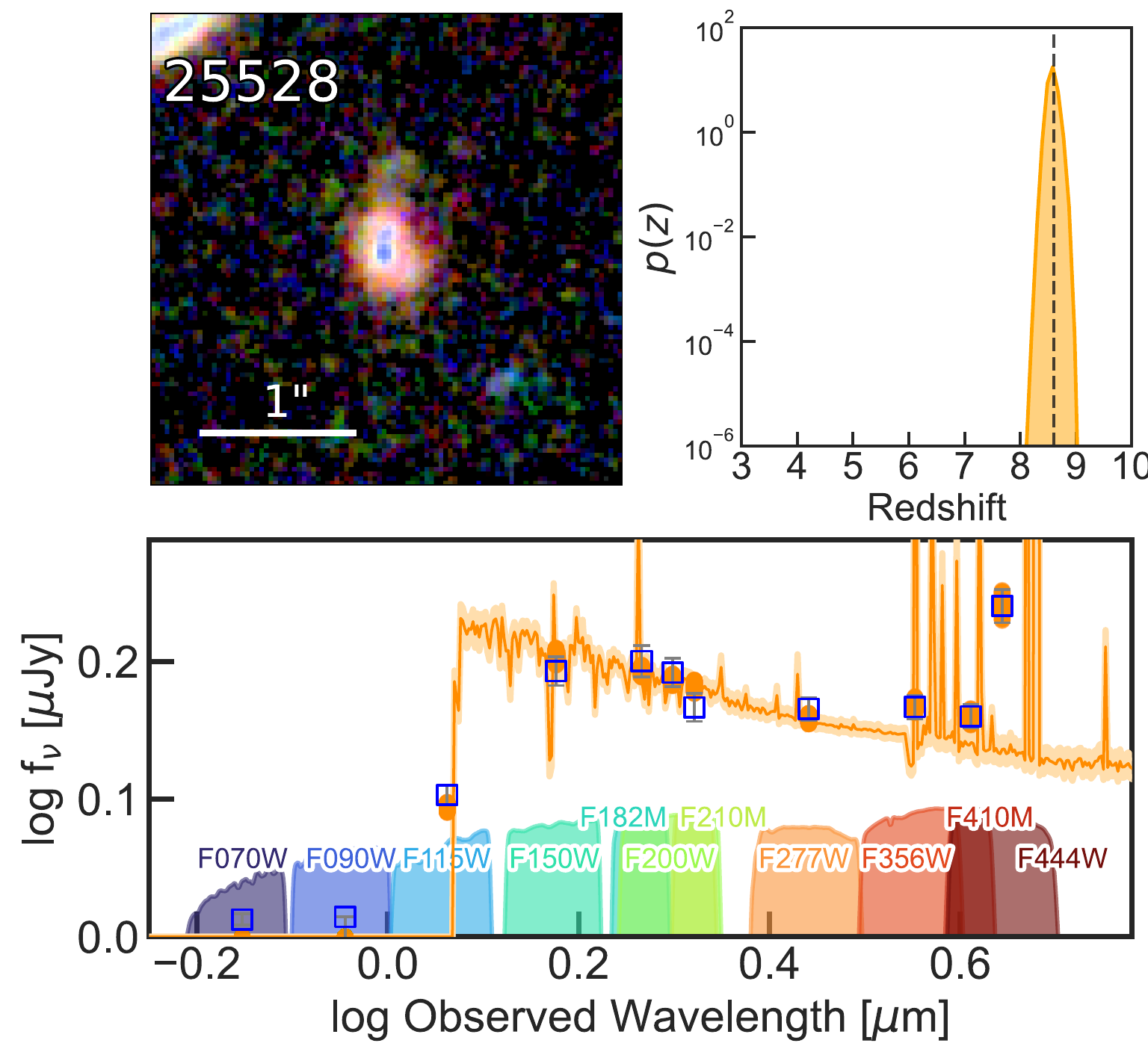}
    \caption{The brightest source in our sample of over-density galaxies, ID-25528. {\bf Upper left panel}: $3\arc\times3\arc$ false color cutout of ID-25528 using F150W, F200W, F444W images. North is up and east is left. {\bf Upper right panel}: Photometric redshift probability of ID-25528. 
    {\bf Lower panel}: Best fit SED for ID-25528.
    ID-25528 is a luminous ($M_{\rm UV}=-21.8\,{\rm mag}$) photometric candidate member galaxy. ID-25528 is spatially resolved to have an elliptical shape with a possible tidal tail toward north, likely indicating recent merger events. The location of ID-25528 is indicated using a white label in Fig. \ref{fig:entire}.
    }
    \label{fig:ID25528-label}
\end{figure}

Among the photometric candidates, ID-25528 (${\rm RA}=63.9968695$, ${\rm Dec}=-24.1184662$) stands out for its brightness (${ m}_{\rm F200W} = 25.7\,{\rm mag}$) and its secure photometric redshift ($z_{\rm ph}=8.60\pm0.1$).
The accuracy of the photometric redshift is supported by the clear Lyman-$\alpha$ break in the F115W image and the enhanced F444W flux, which is fully consistent with the presence of strong \Hb\ and \Oiiiboth\ lines (Figure \ref{fig:ID25528-label}).
If this redshift is confirmed, ID-25528 would rank among the most luminous galaxies at $z>8$ ($M_{\rm UV}=-21.8\,{\rm mag}$) with JWST's spectroscopic confirmation \citep[e.g.,][]{Roberts-Borsani2024,DEugenio2024}.
Unfortunately, our grism observations capture ID-25528 only at the edge of the field of view and only cover a limited wavelength range ($3.1$ -- $4.3\,{\rm \mu m}$).
Thus, no clear emission lines are detected in this wavelength range, as expected from its estimated redshift.
Additionally, a nearby bright foreground galaxy affects its continuum subtraction and elevates the noise in the spectra, preventing detection of other fainter emission lines, such as H$\gamma$ and [O{\sc ii}]$\,\lambda\lambda,3727,3730$ at the shorter wavelength.

Therefore, there is no conclusive evidence that ID-25528 is part of MACS0416-OD-z8p5. However, recent observations suggest that massive star-forming galaxies tend to reside in galaxy over-densities \citep[e.g.,][]{Hashimoto2023, Ma2024}. The apparently large size and tidal tail-like feature of ID-25528 suggest possible recent merger events induced by a dense surrounding environment, such as the paired galaxies ID-17832/ID-17882 and ID1967/ID1988 in the spectroscopic sample.
This feature, thus, makes ID-25528 more likely to be part of MACS0416-OD-z8p5. Follow-up spectroscopy of ID-25528 will be crucial for understanding massive galaxy formation within large-scale galaxy over-densities.

\subsection{Emission Line Velocity and Flux Measurements}

The \Oiiilong\ and \Hb\ line fluxes are measured using extracted one-dimensional (1D) spectra.
During the fitting process, we use a single Gaussian for each line, convolved with the line spread function (LSF) modeled for the NIRCam R and C grisms \citep{Sun2023}. These fitting procedures are performed separately for spectra obtained using the Grism R and C.
The central wavelengths, line widths, and fluxes are left free parameters for the \Oiiilong\ line. For \Oiiishort\ and \Hb, the line widths are fixed to those of \Oiiilong, except for ID-8760, which appears to have larger \Hb\ line width (see \S\ref{sec:AGN} below).
When emission lines are not significantly detected ($<3\,\sigma$), we calculate $3\,\sigma$ flux upper limits by quadratically summing the flux uncertainties of channels over $2\times$ the \Oiiilong\ line FWHM of each galaxy. When \Oiiilong\ line widths are unavailable due to contamination of nearby sources, we assume $150\,{\rm km/s}$.
Measured line fluxes and upper limits are summarized in Table \ref{tab:galaxies}.


\subsection{Spectral Energy Distribution Fitting}
\label{sec:sedfit}

To investigate galaxy properties, we performed spectral energy distribution (SED) fitting using the publicly available code \texttt{BAGPIPES} \citep[][]{Carnall2018} which utilizes the \citet{Bruzual2003} stellar population synthesis models with an initial mass function from \citet{Kroupa2001}.
For the SED fitting, we assumed the following flat priors: a delayed‑tau star‑formation history; a flat prior on ${\rm \tau}$ between $0.1$ and $10\,{\rm Gyr}$; an age ranging from $0.1\,{\rm Myr}$ up to the age of the Universe; a flat metallicity prior ranging from 0.005 to 5.0\,${\rm Z_{\odot}}$; and a formed stellar mass between $10^{5}$ and $10^{15}\,{\rm M_{\odot}}$.
We also adopted the Calzetti dust attenuation law \citep{Calzetti2000} with $A_{\rm v}$ between $0$ and $6\,{\rm mag}$, and assumed a flat prior in logarithmic space for the ionization parameter, with $\log(U)$ ranging from $-1$ to $-4$. The redshifts were fixed to their measured spectroscopic values or the best fit photometric redshifts for photometric samples. In addition to the standard star‑formation rates (SFRs) averaged over $100\,{\rm Myr}$, we computed SFRs averaged over $10\,{\rm Myr}$. The results of the SED fitting are summarized in Table \ref{tab:sedfit} and Table \ref{tab:photoz} for the spectroscopic sample and the photometric sample respectively.

\subsection{Metallicity}

We estimate oxygen abundances ($12+{\rm log(O/H)}$) of spectroscopic member galaxies and candidates using H$\beta$ and \Oiiilong\ lines (i.e., ``Strong-line'' R3 method; \citealt{Kewley2002}). We used the R3 calibration for high-redshift galaxies presented in \citep[][]{Curti2024}. Estimated oxygen abundances ($12+{\rm log(O/H)}$) are listed in Table \ref{tab:sedfit}.

We note that existing metallicity measurements are calibrated using field galaxies and studies suggest that galaxies in over-densities might have evolved, highly metal-enriched properties \citep{Champagne2024a}. Thus, their R3 index calibration might be different from field galaxies. 
The detailed analysis of the mass metallicity relation and its environmental dependence will be further discussed in a separate future study (Hsiao et al., in preparation).


\section{Results} \label{sec:results}

\subsection{Density Enhancement of MACS0416-OD-z8p5}

We estimate the density enhancement of MACS0416-OD-z8p5 based on the definition of $\delta = ({\bar n} - n) / n$, where ${\bar n}$ is the average number density and $n$ is the observed number density of galaxies.

The observed area of SAPPHIRES-EDR is $16.5\,{\rm arcmin^2}$ \citep{Sun2025}. As member galaxies of MACS0416-OD-z8p5 are distributed over the observed area (Fig. \ref{fig:entire}), we assume MACS0416-OD-z8p5 has the same projected area as the SAPPHIRES-EDR observation field.
The separation in transverse comoving distance corresponding to an arcminute at redshift of $8.47$ is $2.63\,{\rm cMpc\,arcmin^{-1}}$, meaning we have $114\,{\rm cMpc^{2}}$ of projected area in the observation.
We then assume that the line of sight comoving size of the over-density is $15\,{\rm cMpc}$ between $z=8.427$ and $z=8.483$, taken from the minimum and maximum redshift of galaxies in the over-density. Finally, we have $1598\,{\rm cMpc^3}$ of volume in the above redshift interval. As we have $7$ -- $9$ galaxies in the MACS0416-OD-z8p5, we now have a calculated galaxy density of $n = 0.0044$ -- $0.0056\,{\rm cMpc^{-3}}$.

To derive the average galaxy number density of ${\bar n}$ at $z\sim8.5$, we base on the UV luminosity function obtained in \citet{Bouwens2021} at $z=8.9$. We integrate the luminosity function down to $M_{\rm UV}=-18$ which is the faint limit of our sample galaxies, and obtain ${\bar n} = 6.4 \times 10^{-4}\,{\rm cMpc^{-3}}$ for the field galaxy density. Finally, using these values, we thus derive $\delta = \frac{n}{{\bar n}} - 1 =5.8\pm1.2$ -- $7.7\pm1.6$, which is significantly larger than the proto-cluster criteria \citep[e.g.,][]{Chiang2013}.  

The derived density enhancement can be a lower limit. In particular, the area evaluating the density enhancement is at least a factor 2 smaller than the areas used for efficient over-density detections in the theoretical prediction \citep{Chiang2013}. Thus, it is not clear whether our observations correspond to the core of the over-density.
Indeed, as predicted by the theoretical work, member galaxies of MACS0416-OD-z8p5 are smoothly distributed over the survey footprint (Figure \ref{fig:entire}), and luminous member candidate (ID-25528) exists close to the edge of the survey footprint (see also \S\ref{sec:lss} for discussion about known galaxies in the existing MACS0416). The SAPPHIRES-EDR fields were observed by the parallel observation strategy. Therefore, sensitive ancillary data are unavailable in the field and searching for potential member galaxies outside of the current foot print is not possible. Wider field observations of nearby fields are essential to provide the full sky distributions of MACS0416-OD-z8p5.

\subsection{Identifying Active Galactic Nuclei}
\label{sec:AGN}

\begin{figure*}[tb]
    \centering
    \includegraphics[width=0.8\textwidth]{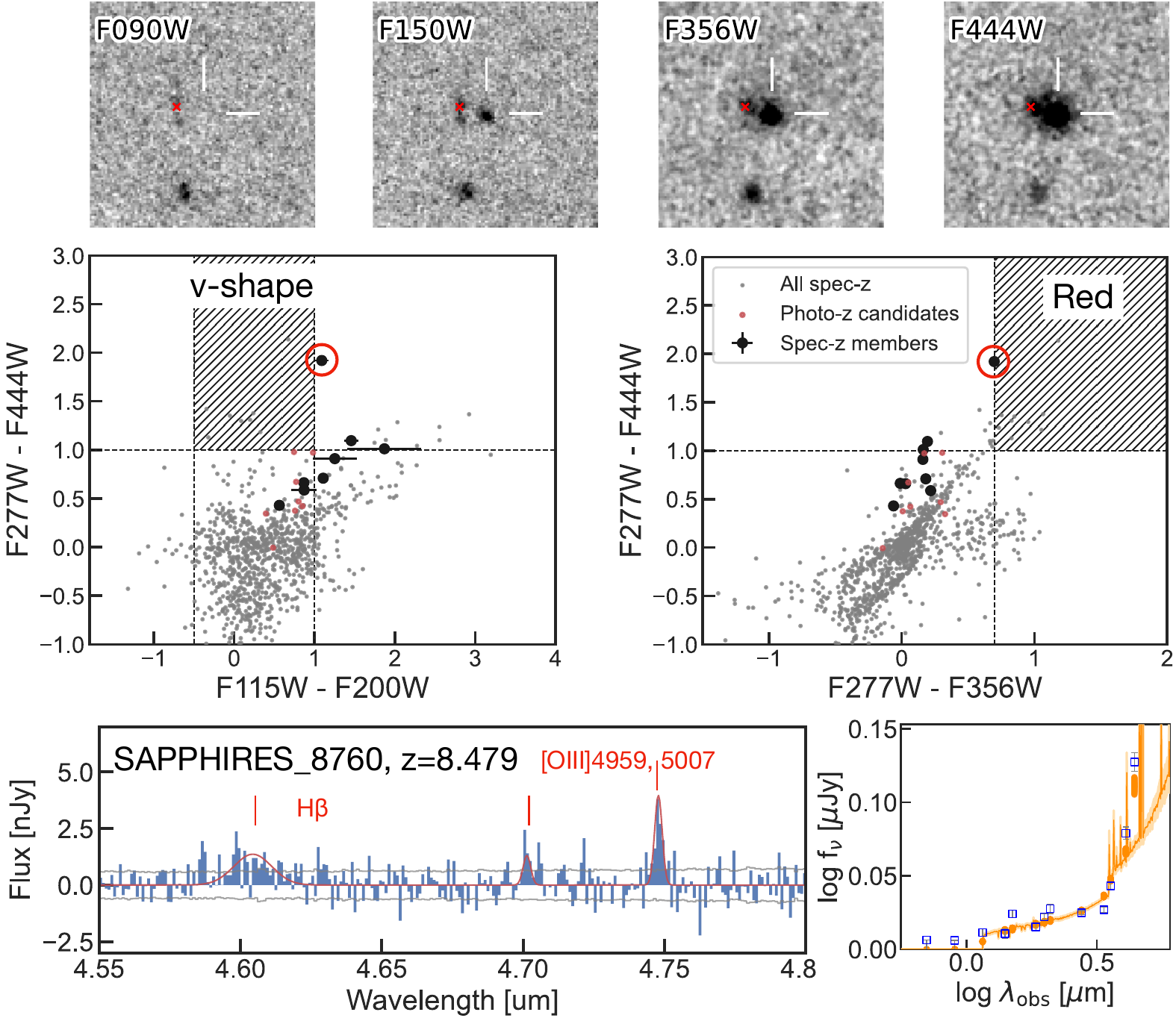}
    \caption{
    {\it Upper panels}: $3\arc\times3\arc$ cutouts of ID-8760 using F090W, F150W, F356W, and F444W filters. Red crosses show a projected foreground galaxy next to the ID-8760. White lines guide the location of ID-8760 in the cutouts.
    {\it Middle panels}: The location of the member galaxies of MACS0416-OD-z8p5 in the color-color plot. The hatched regions show the color-color criteria of LRDs reported in \citet{Greene2024}. Large black circles indicate spectroscopic member galaxies of MACS0416-OD-z8p5, red dots are photometric candidates, and gray dots are spectroscopic samples obtained in SAPPHIRES-EDR.
    {\it Lower panels} shows the grism spectra  and the SED of ID-8760. Gray lines show the $\pm1\,\sigma$ uncertainty of fluxes and the red line shows the result of Gaussian fitting. While there is a close foreground galaxy, morphology shows ID-8760 is very compact and consistent with a point source. The orange SED represents the best-fit SED assuming galaxy only component using \texttt{BAGPIPES}\citep{Carnall2018}.
    For both color criteria, ID-8760 satisfies LRD conditions.  Also, while with a relatively low SNR, the $H\beta$ line shows a broad profile ($\sim1000\,{\rm km/s}$). See also the full SED estimated for ID-8760 in Figure \ref{fig:spectra_summary} in Appendix. All these features show that ID-8760 is an AGN/LRD at $z=8.479$.}
    \label{fig:LRD}
\end{figure*}

JWST has identified a new population of galaxies that are candidates of faint active galactic nuclei (AGNs) \citep{2023ApJ...959...39H,Kocevski2023}. These galaxies are often named as ``Little-Red-Dots''  \citep[LRD;][]{Matthee2024} based on their apparent colors, morphology, and their broad Balmer emission lines.
Despite extensive efforts so far, it is not fully elucidated how and where these LRDs emerge. Furthermore, LRD's physical properties are not yet fully understood.
Theoretical studies suggest an enhanced growth of central black holes of galaxies in dense regions. Recent studies identified several LRDs in a galaxy proto-cluter at high redshift \citep{Champagne2024b,Labbe2024}, although it remains unclear or could be unlikely that LRDs preferentially reside in over-dense environments. 
We examined if enhancement of the AGN population in over-density environments is already the case for this $z\sim8.5$ galaxy over-density by searching for AGN candidates in MACS0416-OD-z8p5.

We used the photometric LRD criterion developed in \citet{Greene2024} and applied the color and morphology classification for the spectroscopically confirmed samples and photometric candidates.
 Among the photometric candidates, we do not find any galaxies satisfying the criteria.
Among the spectroscopic samples, ID-8760 is found to be fully consistent with the color-color criteria of LRD within $\sim1\,\sigma$ (Figure \ref{fig:LRD}).
ID-8760 has $f_{\rm F444W}(0\farcs4)$/$f_{\rm F444W}(0\farcs2)=1.58\pm0.04$, where $f_{\rm F444W}(0\farcs4)$ and $f_{\rm F444W}(0\farcs2)$ represents aperture flux densities measured using aperture diameters of $0\farcs4$ and $0\farcs2$, respectively.
Recently, \citet{Greene2024} used $f_{\rm F444W}(0\farcs4)$/$f_{\rm F444W}(0\farcs2)<1.5$ as a compactness criterion for LRD, which is slightly lower than that for ID-8760 \citep[see also][for a compact criterion of 1.7]{Labbe2025}. However, we identified that there is a foreground galaxy next to the ID-8760, which contributes to elevating the $f_{\rm F444W}(0\farcs4)$. Thus, the current measurement of $f_{\rm F444W}(0\farcs4)$/$f_{\rm F444W}(0\farcs2)=1.58\pm0.04$ is consistent with the compactness criterion of LRDs.
These morphology and color criteria show that ID-8760 is indeed a good candidate to be an LRD and/or AGN.

In the spectrum (Figure \ref{fig:LRD}), the H$\beta$ line of ID-8760 has a large full-width-at-half-maximum of $1010\pm190\,{\rm km/s}$ while the $H\beta$ line is detected with a relatively low SNR (peak SNR of $2.5\,\sigma$). 
Combining the color-color criteria, the compactness, and the large width exceeding $1000\,{\rm km/s}$, we conclude that it is indeed a spectroscopically confirmed LRD/AGN at $z=8.47$. Thus, ID-8740 is among the highest redshift LRDs so far confirmed using {\it JWST} spectroscopy \citep{Kokorev2023, Larson2023}.

\begin{table*}[tb]
    \centering
    \caption{Estimated galaxy properties of spectroscopic samples obtained using BAGPIPES}
    \begin{tabular}{clllccccl}
    \hline
       ID  &  log$(M_{\ast})$ & {\rm SFR$_{10}$} & {\rm SFR$_{100}$} & $M_{\rm UV}$ & $\beta_{\rm UV}$ &  $ A_{\rm V}$ & Age$^{b}$ & $12+{\rm log(O/H)}$ \\
       & ${\rm M_{\odot}}$& ${\rm M_{\odot}/yr^{-1}}$ & ${\rm M_{\odot}/yr^{-1}}$ & ${\rm ABmag}$ & & & Myr &\\
    \hline
1967 & $7.3^{+0.3}_{-0.2}$ & $2.4^{+0.9}_{-0.8}$ & $0.2^{+0.2}_{-0.1}$ & $-18.56\pm0.04$ & $-1.99\pm0.29$& $0.2^{+0.1}_{-0.1}$& $6^{+8}_{-3}$ & $>6.8$\\
1988 & $7.0^{+0.1}_{-0.1}$& $1.1^{+0.3}_{-0.2}$ & $0.10^{+0.03}_{-0.02}$ & $-18.35\pm0.05$ & $-1.96\pm0.97$ & $0.1^{+0.1}_{-0.1}$& $1.7^{+0.9}_{-0.4}$ & $>6.4$\\
8760$^{a}$ & -- & -- & -- & $-19.11\pm0.02$ & $-1.18\pm0.41$ & -- & -- & -- \\
17832 & $7.0^{+0.1}_{-0.1}$ &$1.1^{+0.3}_{-0.2}$& $0.10^{+0.03}_{-0.02}$ & $-18.16\pm0.07$ & $-2.18\pm0.30$ & $0.1^{+0.1}_{-0.1}$& $1.9^{+1.0}_{-0.6}$ & $>7.2$\\
17882 & $7.9^{+0.2}_{-0.2}$ &$17^{+22}_{-10}$ & $0.9^{+0.4}_{-0.3}$ & $-18.46\pm0.06$ & $-3.02\pm1.6$ & $0.5^{+0.2}_{-0.2}$& $1.9^{+0.9}_{-0.6}$ & $>7.1$ \\
17929 & $7.9^{+0.1}_{-0.1}$ &$8.9^{+2.4}_{-1.8}$& $0.9^{+0.2}_{-0.1}$ & $-19.24\pm0.02$ & $-1.92\pm0.40$ & $0.4^{+0.1}_{-0.1}$& $4^{+3}_{-2}$ & $7.62\pm0.03$\\
18743 & $7.6^{+0.2}_{-0.2}$ &$2.6^{+0.7}_{-0.5}$& $0.4^{+0.2}_{-0.1}$ & $-19.02\pm0.02$ & $-1.63\pm0.46$ & $0.1^{+0.1}_{-0.1}$& $11^{+12}_{-5}$ & $>7.2$\\
19406 & $7.9^{+0.1}_{-0.1}$ &$9.7^{+2.1}_{-2.3}$& $0.8 ^{+0.3}_{-0.2}$ & $-20.13\pm0.01$ & $-2.28\pm0.18$ & $0.2^{+0.1}_{-0.1}$& $7^{+5}_{-3}$ & $7.60\pm0.02$\\
21284 & $8.8^{+0.1}_{-0.1}$ &$18.0^{+4.3}_{-3.4}$& $6.7^{+1.6}_{-1.7}$ & $-20.48\pm0.01$ & $-2.02\pm0.19$& $0.3^{+0.1}_{-0.19}$& $70^{+20}_{-20}$ & $>7.6$\\
    \hline
    \multicolumn{8}{c}{Tentative Spectroscopic Sample}\\
    \hline
    16702 & $8.1^{+0.1}_{-0.1}$ & $4.3^{+1.1}_{-1.0}$ & $1.6^{+0.3}_{-0.3}$ & $-18.31\pm0.04$ & $-1.67\pm0.05$ & $0.3^{+0.1}_{-0.1}$ &  $80^{+20}_{-20}$ & $>6.8$\\
    \hline
    \multicolumn{8}{l}{(a) This source is confirmed to be an LRD. SED fitting results are not reliable.}\\
    \multicolumn{8}{l}{(b) Time since star formation began, assuming the delayed-$\tau$ star-formation history}\\
    \end{tabular}
    \label{tab:sedfit}
\end{table*}


\begin{figure}
    \centering
    \includegraphics[width=0.99\columnwidth]{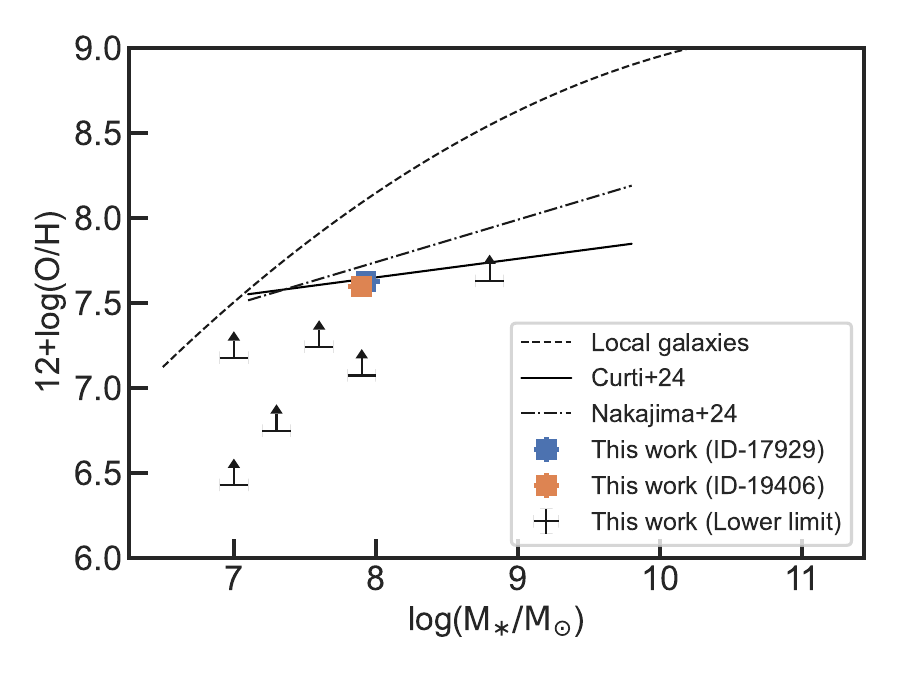}
    \caption{Oxygen abundance as a function of stellar mass for member galaxies of MACS0416-OD-z8p5. Solid line and dash-dotted line show JWST observations of $z>4$ galaxy \citep{Curti2024,Nakajima2023}. The relation for local galaxies is shown with dashed line \citep{Tremonti2004}. Our $3\,\sigma$ lower limits for \Hb\ non-detected galaxies are shown with arrows. Results from two \Hb\ line detected galaxies are shown with squares. While the majority of member galaxies only have lower limits and we only have small number statistics, the \Hb\ detected galaxies show that their metallicity property are consistent with the those of field galaxies at $z>4$.}
    \label{fig:MZR}
\end{figure}

\section{Discussion} \label{sec:discussion}

\subsection{Dark Matter Halo Mass}
Estimating the dark matter halo masses of galaxies is crucial for understanding their formation history, possible descendants, and the theoretical framework underlying their formation history \citep[e.g.,][]{Somerville2015, Wechsler2018}. Additionally, the connection between galaxy and halo formation provides an important constraints on cosmological models \citep[e.g.,][]{Labbe2023, Xiao2024}. Thus, we estimate the dark matter halo mass hosting MACS0416-OD-z8p5 and compare with theoretical expectations.

\subsubsection{Estimation from the observations}

To estimate the dark matter halo masses of member galaxies, we applied two methods to conservatively determine the total dark matter halo mass of MACS0416-OD-z8p5.
We first use the empirically derived $M_{\rm UV}$ -- $M_{\rm halo}$ relation from \citet{Mason2023}.
In particular, we estimate the halo masses of individual galaxies using the $M_{\rm UV}$ -- $M_{\rm halo}$ relation at $z=8$, based on median stellar age with dust attenuation (Figure 3. of \citealt{Mason2023}). The UV magnitudes are derived from F150W fluxes.
By summing the individual halo masses, we obtain a total halo mass for MACS0416-OD-z8p5 of ${\rm log}(M_{\rm halo}/{\rm M_{\odot}})=11.7\pm{0.1}$.
Next, we apply the stellar-to-halo mass ratio from the UniverseMachine simulation \citep{Behroozi2019}. Using the stellar masses obtained in the SED fitting (Table \ref{tab:sedfit}) and the median stellar-to-halo mass ratio at $z=8.4$, we find the total halo mass of ${\rm log}(M_{\rm halo}/{\rm M_{\odot}})=11.4\pm0.1$. 
The uncertainties are estimated by quadratically combining the 16th -- 84th percentile of the stellar-to-halo mass ratio and uncertainties in the stellar masses.
Although small, the difference between both halo mass measurements could be due to the uncertainty in the estimates of  the burstiness of the star-formation activity, which may increase the scatter of the $M_{\rm UV}$ -- $M_{\rm halo}$ relation at high redshift.

\begin{figure*}
    \centering
    \includegraphics[width=0.95\textwidth]{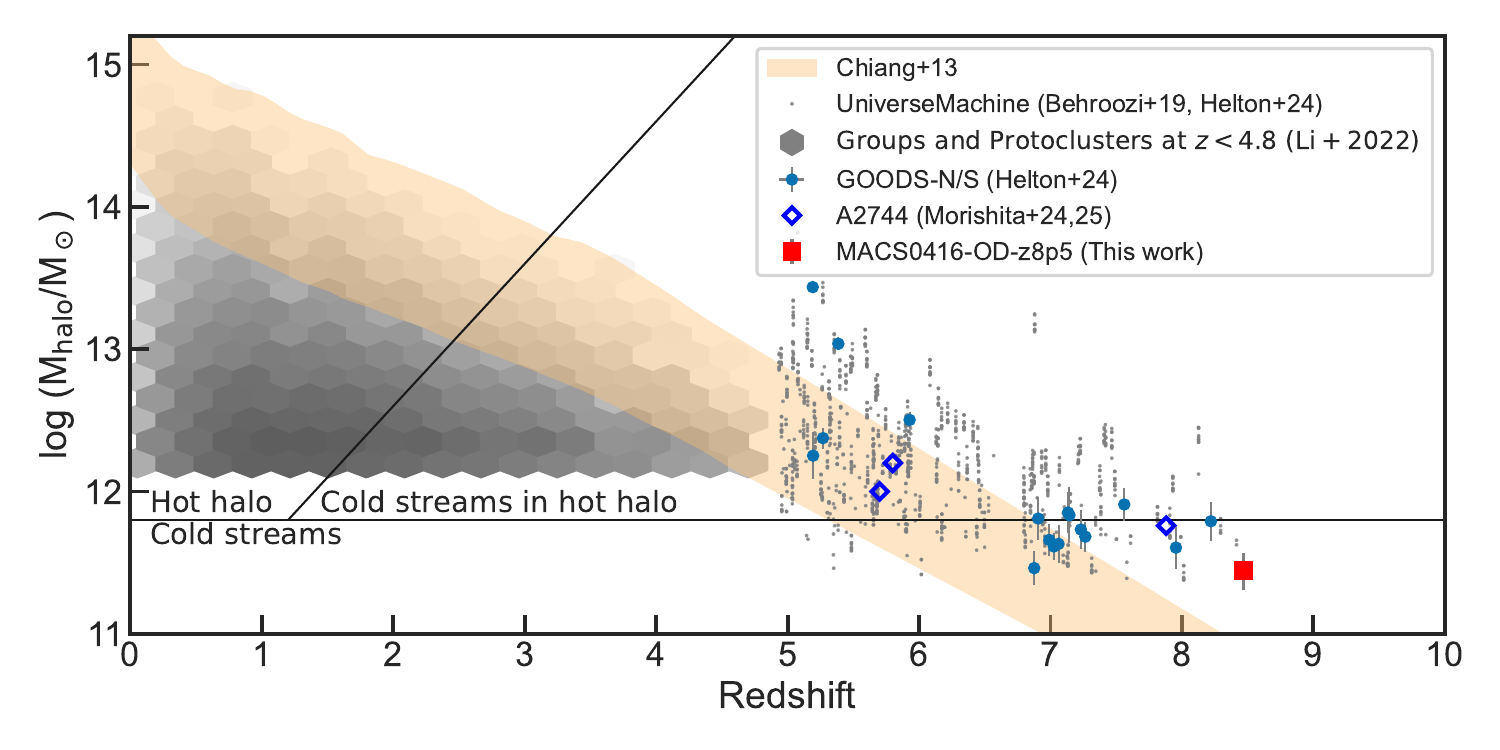}
    \caption{Dark matter halo mass of as a function of redshift. Blue points show data measurements from over-densities in GOODS-N/S \citep{Helton2024}. Blue Diamonds are proto-cluster candidates identified in  \citet{Morishita2024, Morishita2025}. The red square shows our measurements for MACS0416-OD-z8p5. Yellow band shows theoretical predictions of how dark matter halo grows that will become galaxy clusters at $z=0$ \citep{Chiang2013}. Gray-scaled hexagons shows galaxies and proto-clusters at $z\lesssim5$ \citep{Li2022}.
    Gray dots show redshifts and halo masses for theoretical overdensities identified in the UniverseMachine simulations \citep{Helton2024}.
    Solid lines show the predicted transition of the cold stream and shock heated hot halos for each individual halo \citep{Dekel2006}.
    }
    \label{fig:halomass}
\end{figure*}

We note that the estimated halo mass might represent a lower limit, as we are currently missing other massive member galaxies such as a photometric candidate ID-25528 (see \S\ref{sec:photoz}). Inclusion of even a single galaxy like ID-25528 would increase the sum of the halo mass by $0.3\,{\rm dex}$.
Nevertheless, the estimated total halo masses in both methods similarly show that, altogether, the dark matter halo hosting MACS0416-OD-z8p5 is consistent with the most massive halos at $z>8$ (Figure \ref{fig:halomass}).
To be conservative, we adopt a fiducial value of ${\rm log}(M_{\rm halo}/{\rm M_{\odot}})=11.4\pm0.1$ for our following discussions.

Given its large dark matter halo mass, MACS0416-OD-z8p5 would be expected to undergo substantial mass accretion over cosmic time.
In particular, if system like MACS0416-OD-z8p5 experienced the average mass accretion until the present Universe, it could evolve into a massive dark matter halo by the present day, as suggested by theoretical expectations (e.g.,  \citealt{Behroozi2013,Behroozi2019,Chiang2013}).  Thus MACS0416-OD-z8p5 can satisfy the condition to be a proto-cluster. 
However, as shown in \citealt{Angulo2012}, the most massive halos would not always become the most massive clusters at later times.
Therefore, although caution would be needed to interpret the evolutionary path of high-redshift galaxy over-densities, MACS0416-OD-z8p5 would represent one of the highest redshift proto-cluster candidates known to date.

In Figure \ref{fig:halomass}, we compare the estimated halo mass of MACS0416-OD-z8p5 with previously studied galaxy over-densities at $4\lesssim z\lesssim 8$ from \citep{Champagne2024a,Helton2024,Morishita2023,Morishita2024}. We also compare with the theoretical prediction from the UniverseMachine \citep{Behroozi2019} presented in \citet{Helton2024}.
Located at the highest redshift among the spectroscopically confirmed galaxy over-densities, the halo mass of MACS0416-OD-z8p5 is similar to or slightly lower than those 
of previously found over-densities. The trend of having lower halo mass at higher redshift is consistent with the finding from the UniverseMachine, meaning that the currently known theoretical framework can produce the massive dark matter halo such as those hosting MACS0416-OD-z8p5. However, finding such a rare massive dark matter halo in the current small survey footprint ($\sim16.5\,{\rm arcmin^2}$) could be surprising and we will investigate this in the next section.

\subsubsection{Comparison with theoretical expectations}
\label{sec:theoretical}

As discussed above, our estimated halo mass of MACS0416-OD-z8p5 is large, suggesting that the associated over-density is highly biased and represents the highest-redshift, {\bf spectroscopically confirmed} proto-cluster candidate known to date.
However, our survey covers only a small area ($\sim16.5\,{\rm arcmin^{2}}$) and, given the limited field of view, one might not expect to detect such a rare, massive dark matter halo. To investigate the significance of this detection, we compare our finding with theoretical expectations.

To analyze the over-density and compare with theoretical expectations, we extracted $100$ lightcones from the UniverseMachine empirical model \citep{Behroozi2019,Behroozi2020} applied to the public \textit{VSMDPL} simulation (160 Mpc/$h$ size length, $3840^3$ particles with $\sim 10^7$\msun\ mass resolution, \citealt{Planck2016} cosmology; full details in \citealt{Klypin16,RP16}).  We created mock catalogs that have fields of view corresponding to the current SAPPHIRES-EDR observations. The simulated SFRs of the mock catalogs are calculated based on average SFRs over the interval of snapshot (i.e, $19\,{\rm Myrs}$ at $z\sim8.4$) and by taking into account fluctuations originating from uncertainties of observations (i.e., the \texttt{obs\_SFR} column from the UniverseMachine's lightcones).
From each mock catalog, we select galaxies with ${\rm SFR>2.5\,{\rm M_{\odot}\,yr^{-1}}}$ at $z=8$ -- $9$.
The lower limit of the SFR selection corresponds to values when we consider paired-galaxies as single systems (i.e., ID-1967/ID-1988 and ID-17832/ID-17882), and we use the total number of seven galaxies for this comparison.

We then identify any group of galaxies enclosed in the SAPPHIRES-EDR footprint using a redshift bin of $\Delta z=0.05$ as observed in our sample.

From the $100$ simulated mock catalogs, we only find $4$ cases that have more than $7$ star forming galaxies (the number of galaxies in MACS0416-OD-z8p5 with ${\rm SFR > 2.5\,{\rm M_{\odot}\,yr^{-1}}}$ simultaneously located in the SAPPHIRES-EDR footprint. The number of corresponding mock catalogs mildly increases by lowering the SFR threshold; 6 cases for ${\rm SFR > 2.0\,{\rm M_{\odot}\,yr^{-1}}}$ and 7 cases for ${\rm SFR > 1.7\,{\rm M_{\odot}\,yr^{-1}}}$), which however do not change our overall conclusion.
On the other hand, if we further increase the number of observed star-forming galaxies to $8$ (including ID-16702, a tentative spectroscopic candidate) or to $9$ (including ID-25528, a most likely photometric candidate), we only find $3$ and $2$ simulated mock catalogs, respectively, that host over-densities that contain such a large number of galaxies. This means that finding a large galaxy over-density such as MACS0416-OD-z8p5 in a small field of view ($\sim16.5\,{\rm arcmin^2}$) should be a rare event with a nominal estimated probability of $<4$ -- $7\,\%$.

We note that the estimated probability would be an upper limit as the field-of-view and wavelength coverage do not allow us to confirm all candidates that may exist in this field. If we additionally identify more member galaxies in this field (e.g., more than $3$ galaxies) with similar SFRs, the existing model might have difficulty explaining such a galaxy over-density and we would need to revise our current understanding of galaxy formation in early dense environments. 
Future observations pointing at widely separated fields to constrain the variance of galaxy counts in small fields will be essential to accurately constrain the number density of massive galaxy over-densities at high redshift. 
If an over-abundance of galaxy over-densities is confirmed, it suggest that the currently assumed stellar to halo mass ratio for galaxy over-density would be underestimated and that we would need to study its mechanism, such as increasing the baryon conversion efficiency or assuming larger amounts of cold gas inflows.

\subsection{Structure of MACS0416-OD-z8p5}
\label{sec:extent}

\subsubsection{LRD in the MACS0416-OD-z8p5}
\label{sec:AGN}
In MACS0416-OD-z8p5, we identified an LRD and/or AGN candidate (ID-8760). ID-8760 is  located in the south west part of the current footprint (Figure \ref{fig:entire}) where the surface density of galaxies is lower compared with the north east side of the over-density.

\citet{Champagne2024a} similarly find AGNs are located in outer edge regions of galaxy proto-clusters. The authors discussed that the outer parts of proto-clusters are the preferred location for AGNs to emerge due to the large amount of materials funneling down from large structures such as in the cosmic web. \citet{Morishita2023} reported the environmental conditions for the emergence of AGN and find that an AGN/LRD was found to be isolated from other galaxies, possibly avoiding the most dense regions. The authors discussed that the negative feedback would also influence the growth of surrounding galaxies. 
Although interpretations based on the single detection of an LRD in the over-density's isolated area would be uncertain,  our discovery of ID-8760 may be consistent with the scenarios where the emergence of LRD or AGN may have a dependence on environment and through feedback, affect surrounding galaxies.

\subsubsection{Known Galaxy Over-Density Candidate at $z=8.31$ to $z=8.49$ }
\label{sec:lss}
In the gravitational lensing field behind the MACS0416 cluster, multiple massive galaxies are known to exist and they potentially form a galaxy over-density at $z\gtrsim8$. One of the most representative galaxies is MACS0416-Y1 \citep{Bakx2020,Tamura2019,Tamura2023}. Recently, two galaxies were spectroscopically found to exist at a similar redshift (MACS0416-JD at $z=8.34$ and f090d\_018 at $z=8.49$) \citep{Li2023, Ma2024}. Although these galaxies are located away from MACS0416-OD-z8p5 (with a separation of $2.8\,{\rm arcmin}$ or $\sim0.8\,{\rm pMpc}$), the existence of nearby massive galaxies may suggest there exists a large scale structure further connecting these separate fields.
In particular, we find that when we find galaxy over-densities at $z=8$ -- $9$ in the UniverseMachine (see \S\ref{sec:theoretical}), typically the galaxies in over-densities with ${\rm SFR>2\,{\rm M_{\odot}\,yr^{-1}}}$ extend over $\sim2$ -- $4\,{\rm arcmin}$ ($\sim0.5$--$1.1\,{\rm pMpc}$), suggesting that such an extended galaxy over-density may exist.

However detailed studies of the connection between MACS0416-OD-z8p5 and MACS0416-Y1 need further data between the SAPPHIRES-EDR field and the MACS0416 cluster field areas to confirm the large-scale structure.
To further expand the study comparing MACS0416-OD-z8p5 with theoretical expectations (\S\ref{sec:theoretical}), environmental effects of AGN (\S\ref{sec:AGN}), and large scale structure (\S\ref{sec:lss}), further WFSS observations of galaxies at large physical separations are essential as they are outside of the current SAPPHIRES observations.

\section{Conclusions} \label{sec:conclusion}

In this study, we reported the discovery and analyses of a galaxy proto-cluster candidate, MACS0416-OD-z8p5, at $z\sim8.47$. The proto-cluster candidate was identified using JWST NIRCam's WFSS obtained as part of the cycle-3 treasury program SAPPHIRES. From the deep NIRCam imaging and WFSS observations covering a $\sim16.5\,{\rm arcmin^2}$ field around MACS0416 cluster, we found the following results:
\vspace{0.2cm}

\noindent $\bullet$ From a sample of $9$ galaxies with secure spectroscopic redshifts and one galaxy with a tentative spectroscopic redshift, we robustly identify a galaxy overdensity at $z \sim 8.47$, named MACS0416-OD-z8p5.
The high concentration of spectroscopically confirmed galaxies in a small volume allows us to identify the highest redshift spectroscopically confirmed galaxy over-density currently known, with a density enhancement factor of $\delta\sim6$ -- $8$.
Four of the member galaxies are part of two close pair systems, suggesting on-going interactions or major mergers.
The relatively high merger rate, the two interacting system out of eight systems, possibly explain the accelerated galaxy evolution within dense protocluster environments.
\vspace{0.2cm}

\noindent $\bullet$  While the total number of confirmed sources depends on the interpretation of the close-paired galaxies, the density enhancement ($\delta\sim6$--$8$) exceeds predicted conditions for a galaxy proto-cluster \citep{Chiang2013}.
Therefore, we conclude that MACS0416-OD-z8p5 is a candidate system that will become a galaxy cluster in the present day Universe.
\vspace{0.2cm}

\noindent $\bullet$ In MACS0416-OD-z8p5, we confirm that one of its members (ID-8760) is a candidate low mass AGN or ``Little-Red-Dot'' (LRD) at $z=8.47$. ID-8760 meets the color-color selection criteria of LRD from \citet{Greene2024}, although the compactness criterion is marginally satisfied due to the minor contamination from a nearby foreground galaxy. The \Hb\ emission line shows a large line width of ${\rm FWHM}=1010\pm190\,{\rm km/s}$, albeit with a relatively low signal-to-noise ratio (peak SNR of $2.5\,\sigma$). Based on these criteria, ID-8760 is confirmed to be among the highest-redshift LRDs. 
The relatively isolated environment of this LRD within the over-density is in-line with previously identifications of LRDs in proto-clusters, suggesting that environmental effects may influence LRD formation and/or the LRD's negative feedback on formation of surrounding environments.
\vspace{0.2cm}

\noindent $\bullet$ We estimate the total dark matter halo mass of ${\rm log}(M_{\rm halo}/{\rm M_{\odot}})=11.4^{+0.1}_{-0.1}$ using stellar-to-halo-mass ratio from \citet{Behroozi2019}, and ${\rm log}(M_{\rm halo}/{\rm M_{\odot}})=11.7$ from UV luminosity to halo mass relation from \citet{Mason2023}. These estimates suggest the MACS0416-OD-z8p5 is hosted within one of the most massive dark matter halos at $z>8$. 
\vspace{0.2cm}

\noindent $\bullet$ By comparing the number of galaxies with those simulated \citep{Behroozi2019}, we find that the probability of finding such a massive halo in the limited survey area of $16.5\,{\rm arcmin^2}$ is low ($<4$ -- $7\,\%$). If the discrepancy between observations and simulations is confirmed, it may suggest that our current understanding about galaxy evolution in dense galaxy environments needs some revision, such as higher baryon conversion efficiency or larger amounts of gas accretion.
\vspace{0.2cm}

\noindent  $\bullet$ MACS0416-OD-z8p5 is located $\sim5\,{\rm pMpc}$ away from the previously identified massive galaxies and potential galaxy over-density at $z=8.31$ -- $8.49$, discovered behind the lensing cluster (i.e., MACS0416-Y1, JD, f090d\_018; e.g., \citealt{Tamura2019,Li2023,Ma2024}). These findings in a small field suggest the existence of a large scale structure connecting these regions, making this region important to investigate early galaxy formation in dense environments.
\vspace{0.2cm}

The identification of MACS0416-OD-z8p5 demonstrates the power of combining deep multi-wavelength imaging and WFSS, providing a complete sample of line emitters and to search for galaxy proto-clusters at very high redshift. 
Future observations covering larger areas in this field will help characterize the spatial extent and richness of this over-density. Similarly, future NIRSpec observations of the rest-optical emission lines and continuum will constrain the nebular and stellar population properties for the member galaxies alongside representative field galaxies. Furthermore, the completion of the SAPPHIRES observations will enable an evaluation of accurate number densities of galaxy over-densities at high redshifts, an essential study to constrain phenomena that are responsible for the formation of massive galaxy over-density hosted by massive dark matter halos.

\begin{acknowledgments}
We thank Takahiro Morishita and Jaclyn B. Champagne for helpful discussions.

Y.F.\ is supported by JSPS KAKENHI Grant Numbers JP22K21349 and JP23K13149. AJB acknowledges funding from the ``FirstGalaxies'' Advanced Grant from the European Research Council (ERC) under the European Union's Horizon 2020 research and innovation program (Grant agreement No. 789056). C.N.A.W., Y.Z., and ZJ acknowledge support from the NIRCam Development Contract NAS5-02105 from NASA Goddard Space Flight Center to the University of Arizona. FDE acknowledges support by the Science and Technology Facilities Council (STFC), by the ERC through Advanced Grant 695671 ``QUENCH'', and by the UKRI Frontier Research grant RISEandFALL.

This work is based on observations made with the NASA/ESA/CSA James Webb Space Telescope. The data were obtained from the Mikulski Archive for Space Telescopes at the Space Telescope Science Institute, which is operated by the Association of Universities for Research in Astronomy, Inc., under NASA contract NAS 5-03127 for JWST. These observations are associated with program \#6434.
Support for program \#6434 was provided by NASA through a grant from the Space Telescope Science Institute, which is operated by the Association of Universities for Research in Astronomy, Inc., under NASA contract NAS 5-03127. 

\end{acknowledgments}

%

\vspace{5mm}
\facilities{JWST(STIS)}


\software{astropy \citep{2013A&A...558A..33A,2018AJ....156..123A},  
          Cloudy \citep{2013RMxAA..49..137F}, 
          Source Extractor \citep{1996A&AS..117..393B}
          }



\appendix 

\setcounter{figure}{0}
\renewcommand{\thefigure}{\thesection\arabic{figure}}

\setcounter{table}{0}
\renewcommand{\thetable}{\thesection\arabic{table}}

\section{Cutouts and spectra of MACS0416-OD-z8p5 members}
\begin{figure}
    \centering
    \includegraphics[width=0.85\linewidth]{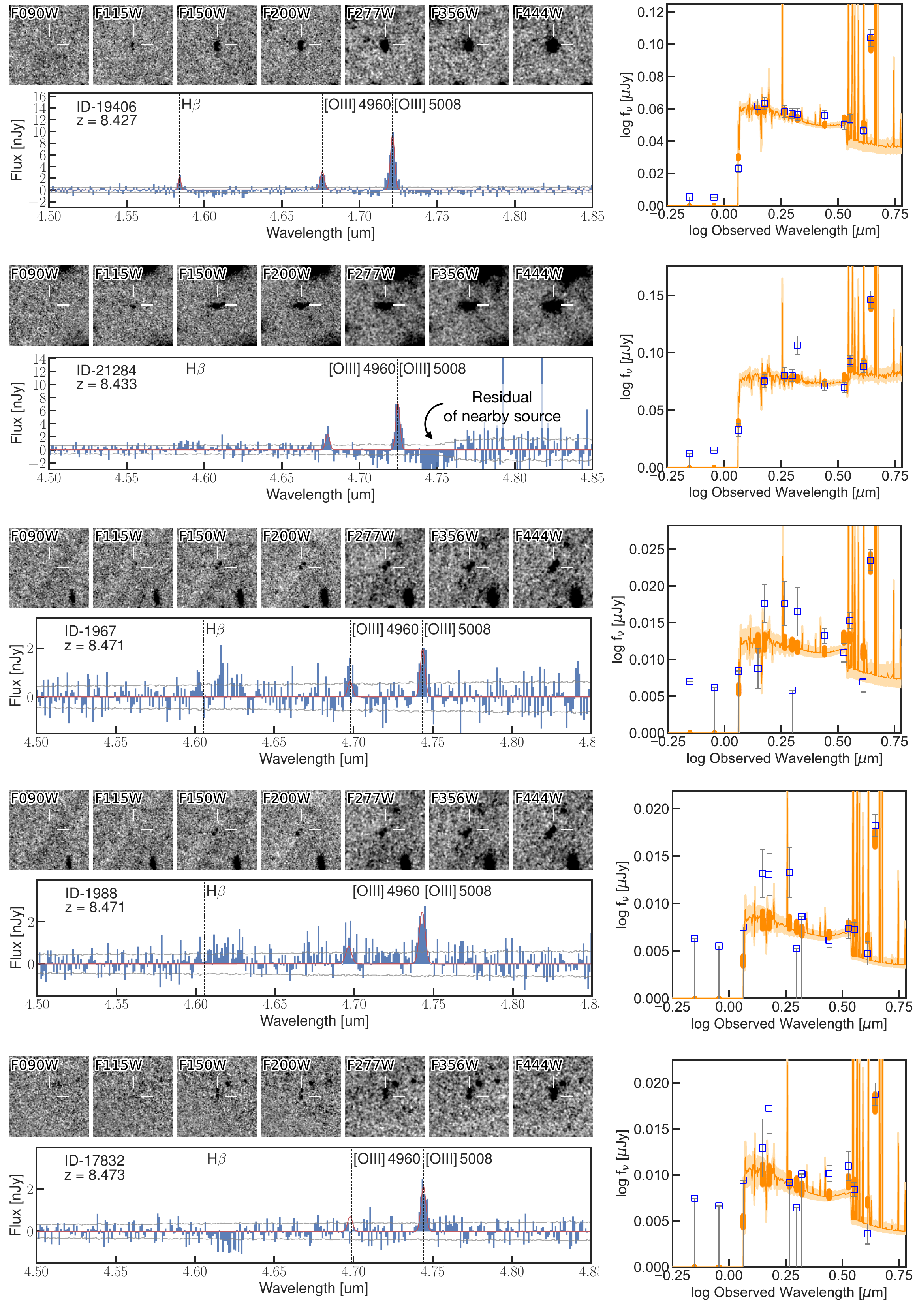}
    \caption{{\it Top left panels}: postage stamp images covering $3\arc\times3\arc$ in each filter used in the SAFFIRES EDR for galaxies belonging to the MACS0416-OD-z8p5 over-density. {\it Bottom left panels:} Extracted spectra using  grism with the F444W filter for member galaxies in MACS0416-OD-z8p5, ordered by their redshifts. Red lines indicate Gaussian fitting results to each emission lines when they are detected with $>3\,\sigma$. Dashed vertical lines show expected wavelengths of \Hb\ and \Oiiiboth\ emission lines based on their measured spectroscopic redshifts.
    {\it right panels}: Estimated SEDs obtained using \texttt{BAGPIPES} \citep{Carnall2018}. Blue squares with errorbars show obtained photometry measured using the Kron apertures. Yellow solid lines and bands represents median and 16th -- 84th percentile distribution of the posterior SEDs.
    }
    \label{fig:spectra_summary}
\end{figure}

In Figure \ref{fig:spectra_summary}, we show spectra of F444W of all spectroscopic member galaxies and the tentative spectroscopic member (ID-16702) of MACS0416-OD-z8p5. 

\setcounter{figure}{0}
\renewcommand{\figurename}{Continued figure}
\begin{figure}
    \centering
    \includegraphics[width=0.85\linewidth]{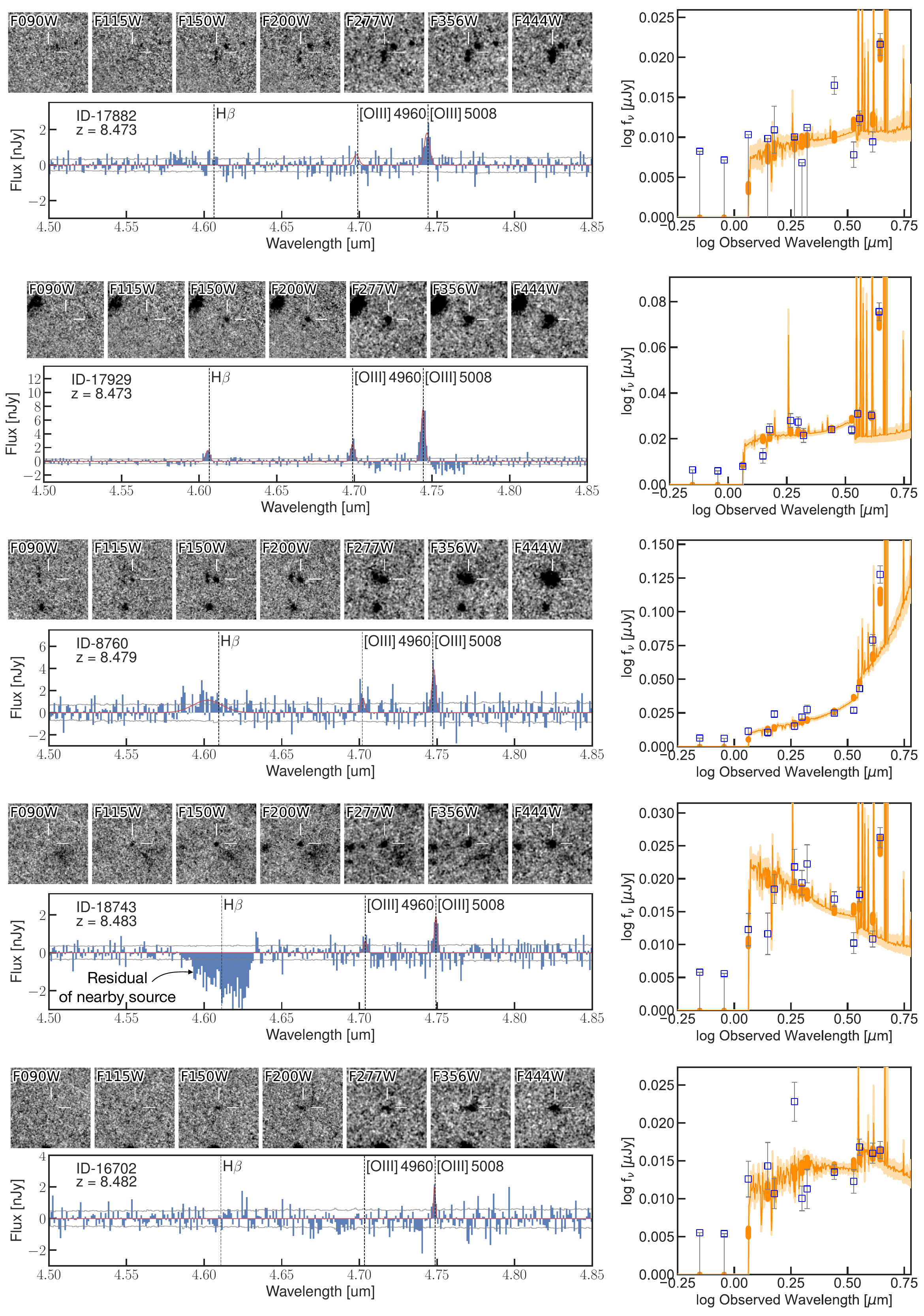}
    \caption{same as the previous figure above.}
    \label{fig:spectra_summary_2}
\end{figure}

\section{Photometric redshift member candidates}

As selected in \S\ref{sec:photoz}, we identified $7$ galaxies that have photometric redshifts consistent with MACS0416-OD-z8p5. In this study, we do not count these photometric candidates as formal member of the over-density. Nevertheless, in Table \ref{tab:photoz}, we list their basic properties for any future follow-up observations and studies. In Table \ref{tab:photoz}, SED fitting results are obtained using the same parameter used in \S\ref{sec:sedfit}, but using their photometric redshifts.

\begin{table*}[h]
    \centering
    \renewcommand{\tabcolsep}{0.12cm}
    \caption{Photometric member candidates selected in \S\ref{sec:photoz} and their  properties.}
    \begin{tabular}{lcccccccccc}
    \hline
       ID  &  RA & Dec &  $z_{\rm phot}$ & ${\rm mag_{F200W}}$ &  ${\rm mag_{F444W}}$ & ${\rm SFR_{10\,Myr}}$ & ${\rm SFR_{100\,Myr}}$ & ${\rm log}(M_{\ast})$ & $A_{\rm V}$ & Age \\
       & deg & deg & & ABmag. & ABmag. & ${\rm M_{\odot}\,yr^{-1}}$ & ${\rm M_{\odot}\,yr^{-1}}$ &    ${\rm M_{\odot}}$ & {\rm mag} & ${\rm Myrs}$ \\
    \hline
    2508 & 63.9499224 & -24.1872654 & $8.56^{+0.25}_{-0.26}$ & $28.89\pm0.06$& $28.37\pm0.02 $ &$2.4^{+1.0}_{-0.7}$ & $0.8^{+0.3}_{-0.2}$ & $7.9^{+0.1}_{-0.2}$ & $0.25^{+0.11}_{-0.09}$ & $0.07^{+0.02}_{-0.02}$\\
    6385 & 63.9597708 & -24.1736440 & $8.56^{+0.35}_{-0.36}$ & $28.87\pm0.08$& $28.93\pm0.04 $ & $2.0^{+0.8}_{-0.5}$& $0.8^{+0.2}_{-0.2}$ & $7.8^{+0.1}_{-0.1}$ & $0.21^{+0.12}_{-0.09}$ & $0.08^{+0.01}_{-0.02}$\\
    8764 & 63.9407058 & -24.1671779 & $8.46^{+0.23}_{-0.33}$ & $29.01\pm0.05$& $28.27\pm0.01 $ & $2.4^{+0.7}_{-0.6}$ & $0.2^{+0.2}_{-0.1}$ & $7.3^{+0.3}_{-0.2}$ & $0.30^{+0.12}_{-0.09}$ & $0.01^{+0.02}_{-0.01}$\\
    15524 & 63.9743115 & -24.1526720 & $8.29^{+0.36}_{-0.30}$ & $28.12\pm0.05$& $27.85\pm0.02 $ & $5.6^{+1.5}_{-1.4}$ & $2.0^{+0.5}_{-0.5}$ & $8.2^{+0.1}_{-0.1}$ & $0.32^{+0.12}_{-0.09}$ & $0.08^{+0.01}_{-0.02}$\\
    16324 & 64.0006973 & -24.1508603 & $8.62^{+0.61}_{-0.46}$& $28.89\pm0.05$& $29.06\pm0.03 $ & $1.1^{+0.4}_{-0.3}$ & $0.4^{+0.1}_{-0.1}$ & $7.5^{+0.1}_{-0.1}$ & $0.10^{+0.10}_{-0.06}$ & $0.07^{+0.02}_{-0.02}$\\
    23481 & 63.9805828 & -24.1297262 & $8.58^{+0.28}_{-0.17}$ & $28.46\pm0.05$& $28.00\pm0.01 $ & $2.9^{+0.7}_{-0.6}$ & $1.0^{+0.3}_{-0.3}$ & $7.9^{+0.1}_{-0.2}$ & $0.31^{+0.11}_{-0.08}$ & $0.07^{+0.02}_{-0.03}$\\
    25528 & 63.9968695 & -24.1184662 & $8.60^{+0.10}_{-0.10}$ & $25.69\pm0.01$& $25.45\pm0.01 $ & $22.9^{+5.3}_{-3.6}$ & $8.3^{+1.8}_{-2.1}$ & $8.9^{+0.1}_{-0.1}$ & $0.07^{+0.05}_{-0.04}$ & $0.07^{+0.02}_{-0.02}$\\
    \hline
    \end{tabular}
    \label{tab:photoz}
\end{table*}


\bibliography{sample631}{}

\begin{thebibliography}{}
\expandafter\ifx\csname natexlab\endcsname\relax\def\natexlab#1{#1}\fi
\providecommand{\url}[1]{\href{#1}{#1}}
\providecommand{\dodoi}[1]{doi:~\href{http://doi.org/#1}{\nolinkurl{#1}}}
\providecommand{\doeprint}[1]{\href{http://ascl.net/#1}{\nolinkurl{http://ascl.net/#1}}}
\providecommand{\doarXiv}[1]{\href{https://arxiv.org/abs/#1}{\nolinkurl{https://arxiv.org/abs/#1}}}

\bibitem[{{Angulo} {et~al.}(2012){Angulo}, {Springel}, {White}, {Jenkins},
  {Baugh}, \& {Frenk}}]{Angulo2012}
{Angulo}, R.~E., {Springel}, V., {White}, S.~D.~M., {et~al.} 2012, \mnras, 426,
  2046, \dodoi{10.1111/j.1365-2966.2012.21830.x}

\bibitem[{{Astropy Collaboration} {et~al.}(2013){Astropy Collaboration},
  {Robitaille}, {Tollerud}, {Greenfield}, {Droettboom}, {Bray}, {Aldcroft},
  {Davis}, {Ginsburg}, {Price-Whelan}, {Kerzendorf}, {Conley}, {Crighton},
  {Barbary}, {Muna}, {Ferguson}, {Grollier}, {Parikh}, {Nair}, {Unther},
  {Deil}, {Woillez}, {Conseil}, {Kramer}, {Turner}, {Singer}, {Fox}, {Weaver},
  {Zabalza}, {Edwards}, {Azalee Bostroem}, {Burke}, {Casey}, {Crawford},
  {Dencheva}, {Ely}, {Jenness}, {Labrie}, {Lim}, {Pierfederici}, {Pontzen},
  {Ptak}, {Refsdal}, {Servillat}, \& {Streicher}}]{2013A&A...558A..33A}
{Astropy Collaboration}, {Robitaille}, T.~P., {Tollerud}, E.~J., {et~al.} 2013,
  \aap, 558, A33, \dodoi{10.1051/0004-6361/201322068}

\bibitem[{{Astropy Collaboration} {et~al.}(2018){Astropy Collaboration},
  {Price-Whelan}, {Sip{\H{o}}cz}, {G{\"u}nther}, {Lim}, {Crawford}, {Conseil},
  {Shupe}, {Craig}, {Dencheva}, {Ginsburg}, {VanderPlas}, {Bradley},
  {P{\'e}rez-Su{\'a}rez}, {de Val-Borro}, {Aldcroft}, {Cruz}, {Robitaille},
  {Tollerud}, {Ardelean}, {Babej}, {Bach}, {Bachetti}, {Bakanov}, {Bamford},
  {Barentsen}, {Barmby}, {Baumbach}, {Berry}, {Biscani}, {Boquien}, {Bostroem},
  {Bouma}, {Brammer}, {Bray}, {Breytenbach}, {Buddelmeijer}, {Burke},
  {Calderone}, {Cano Rodr{\'\i}guez}, {Cara}, {Cardoso}, {Cheedella}, {Copin},
  {Corrales}, {Crichton}, {D'Avella}, {Deil}, {Depagne}, {Dietrich}, {Donath},
  {Droettboom}, {Earl}, {Erben}, {Fabbro}, {Ferreira}, {Finethy}, {Fox},
  {Garrison}, {Gibbons}, {Goldstein}, {Gommers}, {Greco}, {Greenfield},
  {Groener}, {Grollier}, {Hagen}, {Hirst}, {Homeier}, {Horton}, {Hosseinzadeh},
  {Hu}, {Hunkeler}, {Ivezi{\'c}}, {Jain}, {Jenness}, {Kanarek}, {Kendrew},
  {Kern}, {Kerzendorf}, {Khvalko}, {King}, {Kirkby}, {Kulkarni}, {Kumar},
  {Lee}, {Lenz}, {Littlefair}, {Ma}, {Macleod}, {Mastropietro}, {McCully},
  {Montagnac}, {Morris}, {Mueller}, {Mumford}, {Muna}, {Murphy}, {Nelson},
  {Nguyen}, {Ninan}, {N{\"o}the}, {Ogaz}, {Oh}, {Parejko}, {Parley}, {Pascual},
  {Patil}, {Patil}, {Plunkett}, {Prochaska}, {Rastogi}, {Reddy Janga},
  {Sabater}, {Sakurikar}, {Seifert}, {Sherbert}, {Sherwood-Taylor}, {Shih},
  {Sick}, {Silbiger}, {Singanamalla}, {Singer}, {Sladen}, {Sooley},
  {Sornarajah}, {Streicher}, {Teuben}, {Thomas}, {Tremblay}, {Turner},
  {Terr{\'o}n}, {van Kerkwijk}, {de la Vega}, {Watkins}, {Weaver}, {Whitmore},
  {Woillez}, {Zabalza}, \& {Astropy Contributors}}]{2018AJ....156..123A}
{Astropy Collaboration}, {Price-Whelan}, A.~M., {Sip{\H{o}}cz}, B.~M., {et~al.}
  2018, \aj, 156, 123, \dodoi{10.3847/1538-3881/aabc4f}

\bibitem[{{Bakx} {et~al.}(2020){Bakx}, {Tamura}, {Hashimoto}, {Inoue}, {Lee},
  {Mawatari}, {Ota}, {Umehata}, {Zackrisson}, {Hatsukade}, {Kohno}, {Matsuda},
  {Matsuo}, {Okamoto}, {Shibuya}, {Shimizu}, {Taniguchi}, \&
  {Yoshida}}]{Bakx2020}
{Bakx}, T. J.~L.~C., {Tamura}, Y., {Hashimoto}, T., {et~al.} 2020, \mnras, 493,
  4294, \dodoi{10.1093/mnras/staa509}

\bibitem[{{Barrufet} {et~al.}(2023){Barrufet}, {Oesch}, {Weibel}, {Brammer},
  {Bezanson}, {Bouwens}, {Fudamoto}, {Gonzalez}, {Gottumukkala}, {Illingworth},
  {Heintz}, {Holden}, {Labbe}, {Magee}, {Naidu}, {Nelson}, {Stefanon}, {Smit},
  {van Dokkum}, {Weaver}, \& {Williams}}]{Barrufet2023}
{Barrufet}, L., {Oesch}, P.~A., {Weibel}, A., {et~al.} 2023, \mnras, 522, 449,
  \dodoi{10.1093/mnras/stad947}

\bibitem[{{Behroozi} {et~al.}(2019){Behroozi}, {Wechsler}, {Hearin}, \&
  {Conroy}}]{Behroozi2019}
{Behroozi}, P., {Wechsler}, R.~H., {Hearin}, A.~P., \& {Conroy}, C. 2019,
  \mnras, 488, 3143, \dodoi{10.1093/mnras/stz1182}

\bibitem[{{Behroozi} {et~al.}(2020){Behroozi}, {Conroy}, {Wechsler}, {Hearin},
  {Williams}, {Moster}, {Yung}, {Somerville}, {Gottl{\"o}ber}, {Yepes}, \&
  {Endsley}}]{Behroozi2020}
{Behroozi}, P., {Conroy}, C., {Wechsler}, R.~H., {et~al.} 2020, \mnras, 499,
  5702, \dodoi{10.1093/mnras/staa3164}

\bibitem[{{Behroozi} {et~al.}(2013){Behroozi}, {Wechsler}, \&
  {Conroy}}]{Behroozi2013}
{Behroozi}, P.~S., {Wechsler}, R.~H., \& {Conroy}, C. 2013, \apj, 770, 57,
  \dodoi{10.1088/0004-637X/770/1/57}

\bibitem[{{Bertin} \& {Arnouts}(1996)}]{1996A&AS..117..393B}
{Bertin}, E., \& {Arnouts}, S. 1996, \aaps, 117, 393,
  \dodoi{10.1051/aas:1996164}

\bibitem[{{Bouwens} {et~al.}(2021){Bouwens}, {Oesch}, {Stefanon},
  {Illingworth}, {Labb{\'e}}, {Reddy}, {Atek}, {Montes}, {Naidu},
  {Nanayakkara}, {Nelson}, \& {Wilkins}}]{Bouwens2021}
{Bouwens}, R.~J., {Oesch}, P.~A., {Stefanon}, M., {et~al.} 2021, \aj, 162, 47,
  \dodoi{10.3847/1538-3881/abf83e}

\bibitem[{{Brammer} {et~al.}(2008){Brammer}, {van Dokkum}, \&
  {Coppi}}]{Brammer2008}
{Brammer}, G.~B., {van Dokkum}, P.~G., \& {Coppi}, P. 2008, \apj, 686, 1503,
  \dodoi{10.1086/591786}

\bibitem[{{Bromm} \& {Yoshida}(2011)}]{Brom2011}
{Bromm}, V., \& {Yoshida}, N. 2011, \araa, 49, 373,
  \dodoi{10.1146/annurev-astro-081710-102608}

\bibitem[{{Bruzual} \& {Charlot}(2003)}]{Bruzual2003}
{Bruzual}, G., \& {Charlot}, S. 2003, \mnras, 344, 1000,
  \dodoi{10.1046/j.1365-8711.2003.06897.x}

\bibitem[{{Bunker} {et~al.}(2023){Bunker}, {Saxena}, {Cameron}, {Willott},
  {Curtis-Lake}, {Jakobsen}, {Carniani}, {Smit}, {Maiolino}, {Witstok},
  {Curti}, {D'Eugenio}, {Jones}, {Ferruit}, {Arribas}, {Charlot}, {Chevallard},
  {Giardino}, {de Graaff}, {Looser}, {L{\"u}tzgendorf}, {Maseda}, {Rawle},
  {Rix}, {Del Pino}, {Alberts}, {Egami}, {Eisenstein}, {Endsley}, {Hainline},
  {Hausen}, {Johnson}, {Rieke}, {Rieke}, {Robertson}, {Shivaei}, {Stark},
  {Sun}, {Tacchella}, {Tang}, {Williams}, {Willmer}, {Baker}, {Baum},
  {Bhatawdekar}, {Bowler}, {Boyett}, {Chen}, {Circosta}, {Helton}, {Ji},
  {Kumari}, {Lyu}, {Nelson}, {Parlanti}, {Perna}, {Sandles}, {Scholtz},
  {Suess}, {Topping}, {{\"U}bler}, {Wallace}, \& {Whitler}}]{Bunker2023}
{Bunker}, A.~J., {Saxena}, A., {Cameron}, A.~J., {et~al.} 2023, \aap, 677, A88,
  \dodoi{10.1051/0004-6361/202346159}

\bibitem[{{Calzetti} {et~al.}(2000){Calzetti}, {Armus}, {Bohlin}, {Kinney},
  {Koornneef}, \& {Storchi-Bergmann}}]{Calzetti2000}
{Calzetti}, D., {Armus}, L., {Bohlin}, R.~C., {et~al.} 2000, \apj, 533, 682,
  \dodoi{10.1086/308692}

\bibitem[{{Carnall} {et~al.}(2018){Carnall}, {McLure}, {Dunlop}, \&
  {Dav{\'e}}}]{Carnall2018}
{Carnall}, A.~C., {McLure}, R.~J., {Dunlop}, J.~S., \& {Dav{\'e}}, R. 2018,
  \mnras, 480, 4379, \dodoi{10.1093/mnras/sty2169}

\bibitem[{{Carniani} {et~al.}(2024{\natexlab{a}}){Carniani}, {Hainline},
  {D'Eugenio}, {Eisenstein}, {Jakobsen}, {Witstok}, {Johnson}, {Chevallard},
  {Maiolino}, {Helton}, {Willott}, {Robertson}, {Alberts}, {Arribas}, {Baker},
  {Bhatawdekar}, {Boyett}, {Bunker}, {Cameron}, {Cargile}, {Charlot}, {Curti},
  {Curtis-Lake}, {Egami}, {Giardino}, {Isaak}, {Ji}, {Jones}, {Kumari},
  {Maseda}, {Parlanti}, {P{\'e}rez-Gonz{\'a}lez}, {Rawle}, {Rieke}, {Rieke},
  {Del Pino}, {Saxena}, {Scholtz}, {Smit}, {Sun}, {Tacchella}, {{\"U}bler},
  {Venturi}, {Williams}, \& {Willmer}}]{Carniani2024}
{Carniani}, S., {Hainline}, K., {D'Eugenio}, F., {et~al.} 2024{\natexlab{a}},
  \nat, 633, 318, \dodoi{10.1038/s41586-024-07860-9}

\bibitem[{{Carniani} {et~al.}(2024{\natexlab{b}}){Carniani}, {D'Eugenio}, {Ji},
  {Parlanti}, {Scholtz}, {Sun}, {Venturi}, {Bakx}, {Curti}, {Maiolino},
  {Tacchella}, {Zavala}, {Hainline}, {Witstok}, {Johnson}, {Alberts}, {Bunker},
  {Charlot}, {Eisenstein}, {Helton}, {Jakobsen}, {Kumari}, {Robertson},
  {Saxena}, {{\"U}bler}, {Williams}, {Willmer}, \& {Willott}}]{Carniani2024b}
{Carniani}, S., {D'Eugenio}, F., {Ji}, X., {et~al.} 2024{\natexlab{b}}, arXiv
  e-prints, arXiv:2409.20533, \dodoi{10.48550/arXiv.2409.20533}

\bibitem[{{Castellano} {et~al.}(2022){Castellano}, {Fontana}, {Treu},
  {Santini}, {Merlin}, {Leethochawalit}, {Trenti}, {Vanzella}, {Mestric},
  {Bonchi}, {Belfiori}, {Nonino}, {Paris}, {Polenta}, {Roberts-Borsani},
  {Boyett}, {Brada{\v{c}}}, {Calabr{\`o}}, {Glazebrook}, {Grillo}, {Mascia},
  {Mason}, {Mercurio}, {Morishita}, {Nanayakkara}, {Pentericci}, {Rosati},
  {Vulcani}, {Wang}, \& {Yang}}]{Castellano2022}
{Castellano}, M., {Fontana}, A., {Treu}, T., {et~al.} 2022, \apjl, 938, L15,
  \dodoi{10.3847/2041-8213/ac94d0}

\bibitem[{{Champagne} {et~al.}(2024{\natexlab{a}}){Champagne}, {Wang}, {Zhang},
  {Yang}, {Fan}, {Hennawi}, {Sun}, {Ba{\~n}ados}, {Bosman}, {Costa}, {Eilers},
  {Endsley}, {Jin}, {Jun}, {Li}, {Lin}, {Liu}, {Loiacono}, {Lupi},
  {Mazzucchelli}, {Pudoka}, {Protu{\v{s}}ov{\`a}}, {Rojas-Ruiz}, {Tee},
  {Trebitsch}, {Venemans}, {Zhuang}, \& {Zou}}]{Champagne2024a}
{Champagne}, J.~B., {Wang}, F., {Zhang}, H., {et~al.} 2024{\natexlab{a}}, arXiv
  e-prints, arXiv:2410.03826, \dodoi{10.48550/arXiv.2410.03826}

\bibitem[{{Champagne} {et~al.}(2024{\natexlab{b}}){Champagne}, {Wang}, {Yang},
  {Fan}, {Hennawi}, {Sun}, {Ba{\~n}ados}, {Bosman}, {Costa}, {Habouzit}, {Jin},
  {Jun}, {Li}, {Liu}, {Loiacono}, {Lupi}, {Mazzucchelli}, {Pudoka},
  {Rojas-Ruiz}, {Tee}, {Trebitsch}, {Zhang}, {Zhuang}, \&
  {Zou}}]{Champagne2024b}
{Champagne}, J.~B., {Wang}, F., {Yang}, J., {et~al.} 2024{\natexlab{b}}, arXiv
  e-prints, arXiv:2410.03827, \dodoi{10.48550/arXiv.2410.03827}

\bibitem[{{Chiang} {et~al.}(2013){Chiang}, {Overzier}, \&
  {Gebhardt}}]{Chiang2013}
{Chiang}, Y.-K., {Overzier}, R., \& {Gebhardt}, K. 2013, \apj, 779, 127,
  \dodoi{10.1088/0004-637X/779/2/127}

\bibitem[{{Chiang} {et~al.}(2017){Chiang}, {Overzier}, {Gebhardt}, \&
  {Henriques}}]{Chiang2017}
{Chiang}, Y.-K., {Overzier}, R.~A., {Gebhardt}, K., \& {Henriques}, B. 2017,
  \apjl, 844, L23, \dodoi{10.3847/2041-8213/aa7e7b}

\bibitem[{{Curti} {et~al.}(2024){Curti}, {Maiolino}, {Curtis-Lake},
  {Chevallard}, {Carniani}, {D'Eugenio}, {Looser}, {Scholtz}, {Charlot},
  {Cameron}, {{\"U}bler}, {Witstok}, {Boyett}, {Laseter}, {Sandles}, {Arribas},
  {Bunker}, {Giardino}, {Maseda}, {Rawle}, {Rodr{\'\i}guez Del Pino}, {Smit},
  {Willott}, {Eisenstein}, {Hausen}, {Johnson}, {Rieke}, {Robertson},
  {Tacchella}, {Williams}, {Willmer}, {Baker}, {Bhatawdekar}, {Egami},
  {Helton}, {Ji}, {Kumari}, {Perna}, {Shivaei}, \& {Sun}}]{Curti2024}
{Curti}, M., {Maiolino}, R., {Curtis-Lake}, E., {et~al.} 2024, \aap, 684, A75,
  \dodoi{10.1051/0004-6361/202346698}

\bibitem[{{Dayal} {et~al.}(2009){Dayal}, {Ferrara}, {Saro}, {Salvaterra},
  {Borgani}, \& {Tornatore}}]{Dayal2009}
{Dayal}, P., {Ferrara}, A., {Saro}, A., {et~al.} 2009, \mnras, 400, 2000,
  \dodoi{10.1111/j.1365-2966.2009.15593.x}

\bibitem[{{Dekel} \& {Birnboim}(2006)}]{Dekel2006}
{Dekel}, A., \& {Birnboim}, Y. 2006, \mnras, 368, 2,
  \dodoi{10.1111/j.1365-2966.2006.10145.x}

\bibitem[{{D'Eugenio} {et~al.}(2024){D'Eugenio}, {Cameron}, {Scholtz},
  {Carniani}, {Willott}, {Curtis-Lake}, {Bunker}, {Parlanti}, {Maiolino},
  {Willmer}, {Jakobsen}, {Robertson}, {Johnson}, {Tacchella}, {Cargile},
  {Rawle}, {Arribas}, {Chevallard}, {Curti}, {Egami}, {Eisenstein}, {Kumari},
  {Looser}, {Rieke}, {Rodr{\'\i}guez Del Pino}, {Saxena}, {{\"U}bler},
  {Venturi}, {Witstok}, {Baker}, {Bhatawdekar}, {Bonaventura}, {Boyett},
  {Charlot}, {Danhaive}, {Hainline}, {Hausen}, {Helton}, {Ji}, {Ji}, {Jones},
  {Joud{\v{z}}balis}, {Maseda}, {P{\'e}rez-Gonz{\'a}lez}, {Perna},
  {Pusk{\'a}s}, {Shivaei}, {Silcock}, {Simmonds}, {Smit}, {Sun}, {Villanueva},
  {Williams}, \& {Zhu}}]{DEugenio2024}
{D'Eugenio}, F., {Cameron}, A.~J., {Scholtz}, J., {et~al.} 2024, arXiv
  e-prints, arXiv:2404.06531, \dodoi{10.48550/arXiv.2404.06531}

\bibitem[{{Donnan} {et~al.}(2023){Donnan}, {McLeod}, {Dunlop}, {McLure},
  {Carnall}, {Begley}, {Cullen}, {Hamadouche}, {Bowler}, {Magee}, {McCracken},
  {Milvang-Jensen}, {Moneti}, \& {Targett}}]{Donnan2023}
{Donnan}, C.~T., {McLeod}, D.~J., {Dunlop}, J.~S., {et~al.} 2023, \mnras, 518,
  6011, \dodoi{10.1093/mnras/stac3472}

\bibitem[{{Eilers} {et~al.}(2024){Eilers}, {Mackenzie}, {Pizzati}, {Matthee},
  {Hennawi}, {Zhang}, {Bordoloi}, {Kashino}, {Lilly}, {Naidu}, {Simcoe}, {Yue},
  {Frenk}, {Helly}, {Schaller}, \& {Schaye}}]{Eilers2024}
{Eilers}, A.-C., {Mackenzie}, R., {Pizzati}, E., {et~al.} 2024, \apj, 974, 275,
  \dodoi{10.3847/1538-4357/ad778b}

\bibitem[{{Endsley} {et~al.}(2021){Endsley}, {Stark}, {Charlot}, {Chevallard},
  {Robertson}, {Bouwens}, \& {Stefanon}}]{Endsley2021}
{Endsley}, R., {Stark}, D.~P., {Charlot}, S., {et~al.} 2021, \mnras, 502, 6044,
  \dodoi{10.1093/mnras/stab432}

\bibitem[{{Fan} {et~al.}(2006){Fan}, {Strauss}, {Becker}, {White}, {Gunn},
  {Knapp}, {Richards}, {Schneider}, {Brinkmann}, \& {Fukugita}}]{Fan2006}
{Fan}, X., {Strauss}, M.~A., {Becker}, R.~H., {et~al.} 2006, \aj, 132, 117,
  \dodoi{10.1086/504836}

\bibitem[{{Ferland} {et~al.}(2013){Ferland}, {Porter}, {van Hoof}, {Williams},
  {Abel}, {Lykins}, {Shaw}, {Henney}, \& {Stancil}}]{2013RMxAA..49..137F}
{Ferland}, G.~J., {Porter}, R.~L., {van Hoof}, P.~A.~M., {et~al.} 2013, \rmxaa,
  49, 137.
\newblock \doarXiv{1302.4485}

\bibitem[{{Finkelstein} {et~al.}(2022){Finkelstein}, {Bagley}, {Arrabal Haro},
  {Dickinson}, {Ferguson}, {Kartaltepe}, {Papovich}, {Burgarella}, {Kocevski},
  {Huertas-Company}, {Iyer}, {Koekemoer}, {Larson}, {P{\'e}rez-Gonz{\'a}lez},
  {Rose}, {Tacchella}, {Wilkins}, {Chworowsky}, {Medrano}, {Morales},
  {Somerville}, {Yung}, {Fontana}, {Giavalisco}, {Grazian}, {Grogin}, {Kewley},
  {Kirkpatrick}, {Kurczynski}, {Lotz}, {Pentericci}, {Pirzkal}, {Ravindranath},
  {Ryan}, {Trump}, {Yang}, {Almaini}, {Amor{\'\i}n}, {Annunziatella},
  {Backhaus}, {Barro}, {Behroozi}, {Bell}, {Bhatawdekar}, {Bisigello}, {Bromm},
  {Buat}, {Buitrago}, {Calabr{\`o}}, {Casey}, {Castellano}, {Ch{\'a}vez Ortiz},
  {Ciesla}, {Cleri}, {Cohen}, {Cole}, {Cooke}, {Cooper}, {Cooray}, {Costantin},
  {Cox}, {Croton}, {Daddi}, {Dav{\'e}}, {de La Vega}, {Dekel}, {Elbaz},
  {Estrada-Carpenter}, {Faber}, {Fern{\'a}ndez}, {Finkelstein}, {Freundlich},
  {Fujimoto}, {Garc{\'\i}a-Argum{\'a}nez}, {Gardner}, {Gawiser},
  {G{\'o}mez-Guijarro}, {Guo}, {Hamblin}, {Hamilton}, {Hathi}, {Holwerda},
  {Hirschmann}, {Hutchison}, {Jaskot}, {Jha}, {Jogee}, {Juneau}, {Jung},
  {Kassin}, {Le Bail}, {Leung}, {Lucas}, {Magnelli}, {Mantha}, {Matharu},
  {McGrath}, {McIntosh}, {Merlin}, {Mobasher}, {Newman}, {Nicholls}, {Pandya},
  {Rafelski}, {Ronayne}, {Santini}, {Seill{\'e}}, {Shah}, {Shen}, {Simons},
  {Snyder}, {Stanway}, {Straughn}, {Teplitz}, {Vanderhoof}, {Vega-Ferrero},
  {Wang}, {Weiner}, {Willmer}, {Wuyts}, {Zavala}, \& {Ceers
  Team}}]{Finkelstein2022}
{Finkelstein}, S.~L., {Bagley}, M.~B., {Arrabal Haro}, P., {et~al.} 2022,
  \apjl, 940, L55, \dodoi{10.3847/2041-8213/ac966e}

\bibitem[{{Fudamoto} {et~al.}(2021){Fudamoto}, {Oesch}, {Schouws}, {Stefanon},
  {Smit}, {Bouwens}, {Bowler}, {Endsley}, {Gonzalez}, {Inami}, {Labbe},
  {Stark}, {Aravena}, {Barrufet}, {da Cunha}, {Dayal}, {Ferrara}, {Graziani},
  {Hodge}, {Hutter}, {Li}, {De Looze}, {Nanayakkara}, {Pallottini}, {Riechers},
  {Schneider}, {Ucci}, {van der Werf}, \& {White}}]{Fudamoto2021}
{Fudamoto}, Y., {Oesch}, P.~A., {Schouws}, S., {et~al.} 2021, \nat, 597, 489,
  \dodoi{10.1038/s41586-021-03846-z}

\bibitem[{{Greene} {et~al.}(2024){Greene}, {Labbe}, {Goulding}, {Furtak},
  {Chemerynska}, {Kokorev}, {Dayal}, {Volonteri}, {Williams}, {Wang}, {Setton},
  {Burgasser}, {Bezanson}, {Atek}, {Brammer}, {Cutler}, {Feldmann}, {Fujimoto},
  {Glazebrook}, {de Graaff}, {Khullar}, {Leja}, {Marchesini}, {Maseda},
  {Matthee}, {Miller}, {Naidu}, {Nanayakkara}, {Oesch}, {Pan}, {Papovich},
  {Price}, {van Dokkum}, {Weaver}, {Whitaker}, \& {Zitrin}}]{Greene2024}
{Greene}, J.~E., {Labbe}, I., {Goulding}, A.~D., {et~al.} 2024, \apj, 964, 39,
  \dodoi{10.3847/1538-4357/ad1e5f}

\bibitem[{{Hainline} {et~al.}(2024){Hainline}, {Johnson}, {Robertson},
  {Tacchella}, {Helton}, {Sun}, {Eisenstein}, {Simmonds}, {Topping}, {Whitler},
  {Willmer}, {Rieke}, {Suess}, {Hviding}, {Cameron}, {Alberts}, {Baker},
  {Baum}, {Bhatawdekar}, {Bonaventura}, {Boyett}, {Bunker}, {Carniani},
  {Charlot}, {Chevallard}, {Chen}, {Curti}, {Curtis-Lake}, {D'Eugenio},
  {Egami}, {Endsley}, {Hausen}, {Ji}, {Looser}, {Lyu}, {Maiolino}, {Nelson},
  {Pusk{\'a}s}, {Rawle}, {Sandles}, {Saxena}, {Smit}, {Stark}, {Williams},
  {Willott}, \& {Witstok}}]{Hainline2024}
{Hainline}, K.~N., {Johnson}, B.~D., {Robertson}, B., {et~al.} 2024, \apj, 964,
  71, \dodoi{10.3847/1538-4357/ad1ee4}

\bibitem[{{Harikane} {et~al.}(2019){Harikane}, {Ouchi}, {Ono}, {Fujimoto},
  {Donevski}, {Shibuya}, {Faisst}, {Goto}, {Hatsukade}, {Kashikawa}, {Kohno},
  {Hashimoto}, {Higuchi}, {Inoue}, {Lin}, {Martin}, {Overzier}, {Smail},
  {Toshikawa}, {Umehata}, {Ao}, {Chapman}, {Clements}, {Im}, {Jing},
  {Kawaguchi}, {Lee}, {Lee}, {Lin}, {Matsuoka}, {Marinello}, {Nagao},
  {Onodera}, {Toft}, \& {Wang}}]{Harikane2019}
{Harikane}, Y., {Ouchi}, M., {Ono}, Y., {et~al.} 2019, \apj, 883, 142,
  \dodoi{10.3847/1538-4357/ab2cd5}

\bibitem[{{Harikane} {et~al.}(2023{\natexlab{a}}){Harikane}, {Ouchi}, {Oguri},
  {Ono}, {Nakajima}, {Isobe}, {Umeda}, {Mawatari}, \& {Zhang}}]{Harikane2023}
{Harikane}, Y., {Ouchi}, M., {Oguri}, M., {et~al.} 2023{\natexlab{a}}, \apjs,
  265, 5, \dodoi{10.3847/1538-4365/acaaa9}

\bibitem[{{Harikane} {et~al.}(2023{\natexlab{b}}){Harikane}, {Zhang},
  {Nakajima}, {Ouchi}, {Isobe}, {Ono}, {Hatano}, {Xu}, \&
  {Umeda}}]{2023ApJ...959...39H}
{Harikane}, Y., {Zhang}, Y., {Nakajima}, K., {et~al.} 2023{\natexlab{b}}, \apj,
  959, 39, \dodoi{10.3847/1538-4357/ad029e}

\bibitem[{{Hashimoto} {et~al.}(2023){Hashimoto}, {{\'A}lvarez-M{\'a}rquez},
  {Fudamoto}, {Colina}, {Inoue}, {Nakazato}, {Ceverino}, {Yoshida},
  {Costantin}, {Sugahara}, {G{\'o}mez}, {Blanco-Prieto}, {Mawatari}, {Arribas},
  {Marques-Chaves}, {Pereira-Santaella}, {Bakx}, {Hagimoto}, {Hashigaya},
  {Matsuo}, {Tamura}, {Usui}, \& {Ren}}]{Hashimoto2023}
{Hashimoto}, T., {{\'A}lvarez-M{\'a}rquez}, J., {Fudamoto}, Y., {et~al.} 2023,
  \apjl, 955, L2, \dodoi{10.3847/2041-8213/acf57c}

\bibitem[{{Helton} {et~al.}(2024{\natexlab{a}}){Helton}, {Sun}, {Woodrum},
  {Hainline}, {Willmer}, {Rieke}, {Rieke}, {Alberts}, {Eisenstein},
  {Tacchella}, {Robertson}, {Johnson}, {Baker}, {Bhatawdekar}, {Bunker},
  {Chen}, {Egami}, {Ji}, {Maiolino}, {Willott}, \& {Witstok}}]{Helton2024}
{Helton}, J.~M., {Sun}, F., {Woodrum}, C., {et~al.} 2024{\natexlab{a}}, \apj,
  974, 41, \dodoi{10.3847/1538-4357/ad6867}

\bibitem[{{Helton} {et~al.}(2024{\natexlab{b}}){Helton}, {Sun}, {Woodrum},
  {Hainline}, {Willmer}, {Rieke}, {Rieke}, {Tacchella}, {Robertson}, {Johnson},
  {Alberts}, {Eisenstein}, {Hausen}, {Bonaventura}, {Bunker}, {Charlot},
  {Curti}, {Curtis-Lake}, {Looser}, {Maiolino}, {Willott}, {Witstok}, {Boyett},
  {Chen}, {Egami}, {Endsley}, {Hviding}, {Jaffe}, {Ji}, {Lyu}, \&
  {Sandles}}]{Helton2024b}
---. 2024{\natexlab{b}}, \apj, 962, 124, \dodoi{10.3847/1538-4357/ad0da7}

\bibitem[{{Herard-Demanche} {et~al.}(2025){Herard-Demanche}, {Bouwens},
  {Oesch}, {Naidu}, {Decarli}, {Nelson}, {Brammer}, {Weibel}, {Xiao},
  {Stefanon}, {Walter}, {Matthee}, {Meyer}, {Wuyts}, {Reddy}, {Rowland}, {van
  Leeuwen}, {Haro}, {Dannerbauer}, {Shapley}, {Chisholm}, {van Dokkum},
  {Labbe}, {Illingworth}, {Schaerer}, \& {Shivaei}}]{Herard-Demanche2025}
{Herard-Demanche}, T., {Bouwens}, R.~J., {Oesch}, P.~A., {et~al.} 2025, \mnras,
  \dodoi{10.1093/mnras/staf030}

\bibitem[{{Higuchi} {et~al.}(2019){Higuchi}, {Ouchi}, {Ono}, {Shibuya},
  {Toshikawa}, {Harikane}, {Kojima}, {Chiang}, {Egami}, {Kashikawa},
  {Overzier}, {Konno}, {Inoue}, {Hasegawa}, {Fujimoto}, {Goto}, {Ishikawa},
  {Ito}, {Komiyama}, \& {Tanaka}}]{Higuchi2019}
{Higuchi}, R., {Ouchi}, M., {Ono}, Y., {et~al.} 2019, \apj, 879, 28,
  \dodoi{10.3847/1538-4357/ab2192}

\bibitem[{{Hoag} {et~al.}(2019){Hoag}, {Brada{\v{c}}}, {Huang}, {Mason},
  {Treu}, {Schmidt}, {Trenti}, {Strait}, {Lemaux}, {Finney}, \&
  {Paddock}}]{Hoag2019}
{Hoag}, A., {Brada{\v{c}}}, M., {Huang}, K., {et~al.} 2019, \apj, 878, 12,
  \dodoi{10.3847/1538-4357/ab1de7}

\bibitem[{{Hsiao} {et~al.}(2023){Hsiao}, {Coe}, {Abdurro'uf}, {Whitler},
  {Jung}, {Khullar}, {Meena}, {Dayal}, {Barrow}, {Santos-Olmsted}, {Casselman},
  {Vanzella}, {Nonino}, {Jim{\'e}nez-Teja}, {Oguri}, {Stark}, {Furtak},
  {Zitrin}, {Adamo}, {Brammer}, {Bradley}, {Diego}, {Zackrisson},
  {Finkelstein}, {Windhorst}, {Bhatawdekar}, {Hutchison}, {Broadhurst},
  {Dimauro}, {Andrade-Santos}, {Eldridge}, {Acebron}, {Avila}, {Bayliss},
  {Ben{\'\i}tez}, {Binggeli}, {Bolan}, {Brada{\v{c}}}, {Carnall}, {Conselice},
  {Donahue}, {Frye}, {Fujimoto}, {Henry}, {James}, {Kassin}, {Kewley},
  {Larson}, {Lauer}, {Law}, {Mahler}, {Mainali}, {McCandliss}, {Nicholls},
  {Pirzkal}, {Postman}, {Rigby}, {Ryan}, {Senchyna}, {Sharon}, {Shimizu},
  {Strait}, {Tang}, {Trenti}, {Vikaeus}, \& {Welch}}]{Hsiao2023}
{Hsiao}, T. Y.-Y., {Coe}, D., {Abdurro'uf}, {et~al.} 2023, \apjl, 949, L34,
  \dodoi{10.3847/2041-8213/acc94b}

\bibitem[{{Hsiao} {et~al.}(2024){Hsiao}, {{\'A}lvarez-M{\'a}rquez}, {Coe},
  {Crespo G{\'o}mez}, {Abdurro'uf}, {Dayal}, {Larson}, {Bik}, {Blanco-Prieto},
  {Colina}, {P{\'e}rez-Gonz{\'a}lez}, {Costantin}, {Prieto-Jim{\'e}nez},
  {Adamo}, {Bradley}, {Conselice}, {Fujimoto}, {Furtak}, {Hutchison}, {James},
  {Jim{\'e}nez-Teja}, {Jung}, {Kokorev}, {Mingozzi}, {Norman}, {Ricotti},
  {Rigby}, {Sharon}, {Vanzella}, {Welch}, {Xu}, {Zackrisson}, \&
  {Zitrin}}]{Hsiao2024}
{Hsiao}, T. Y.-Y., {{\'A}lvarez-M{\'a}rquez}, J., {Coe}, D., {et~al.} 2024,
  \apj, 973, 81, \dodoi{10.3847/1538-4357/ad6562}

\bibitem[{{Jin} {et~al.}(2023){Jin}, {Sillassen}, {Magdis}, {Vijayan},
  {Brammer}, {Kokorev}, {Weaver}, {Gobat}, {Gim{\'e}nez-Arteaga}, {Valentino},
  {Brinch}, {G{\'o}mez-Guijarro}, {Shuntov}, {Toft}, {Greve}, \& {Blanquez
  Sese}}]{Jin2023}
{Jin}, S., {Sillassen}, N.~B., {Magdis}, G.~E., {et~al.} 2023, \aap, 670, L11,
  \dodoi{10.1051/0004-6361/202245724}

\bibitem[{{Jin} {et~al.}(2024){Jin}, {Yang}, {Fan}, {Wang}, {Kakiichi},
  {Meyer}, {Becker}, {Zou}, {Ba{\~n}ados}, {Champagne}, {D'Odorico}, {Yue},
  {Bosman}, {Cai}, {Eilers}, {Hennawi}, {Jun}, {Li}, {Li}, {Liu}, {Pudoka},
  {Satyavolu}, {Sun}, {Tee}, \& {Wu}}]{Jin2024}
{Jin}, X., {Yang}, J., {Fan}, X., {et~al.} 2024, \apj, 976, 93,
  \dodoi{10.3847/1538-4357/ad82de}

\bibitem[{{Kewley} \& {Dopita}(2002)}]{Kewley2002}
{Kewley}, L.~J., \& {Dopita}, M.~A. 2002, \apjs, 142, 35,
  \dodoi{10.1086/341326}

\bibitem[{{Klypin} {et~al.}(2016){Klypin}, {Yepes}, {Gottl{\"o}ber}, {Prada},
  \& {He{\ss}}}]{Klypin16}
{Klypin}, A., {Yepes}, G., {Gottl{\"o}ber}, S., {Prada}, F., \& {He{\ss}}, S.
  2016, \mnras, 457, 4340, \dodoi{10.1093/mnras/stw248}

\bibitem[{{Kocevski} {et~al.}(2023){Kocevski}, {Onoue}, {Inayoshi}, {Trump},
  {Arrabal Haro}, {Grazian}, {Dickinson}, {Finkelstein}, {Kartaltepe},
  {Hirschmann}, {Aird}, {Holwerda}, {Fujimoto}, {Juneau}, {Amor{\'\i}n},
  {Backhaus}, {Bagley}, {Barro}, {Bell}, {Bisigello}, {Calabr{\`o}}, {Cleri},
  {Cooper}, {Ding}, {Grogin}, {Ho}, {Hutchison}, {Inoue}, {Jiang}, {Jones},
  {Koekemoer}, {Li}, {Li}, {McGrath}, {Molina}, {Papovich},
  {P{\'e}rez-Gonz{\'a}lez}, {Pirzkal}, {Wilkins}, {Yang}, \&
  {Yung}}]{Kocevski2023}
{Kocevski}, D.~D., {Onoue}, M., {Inayoshi}, K., {et~al.} 2023, \apjl, 954, L4,
  \dodoi{10.3847/2041-8213/ace5a0}

\bibitem[{{Kokorev} {et~al.}(2023){Kokorev}, {Fujimoto}, {Labbe}, {Greene},
  {Bezanson}, {Dayal}, {Nelson}, {Atek}, {Brammer}, {Caputi}, {Chemerynska},
  {Cutler}, {Feldmann}, {Fudamoto}, {Furtak}, {Goulding}, {de Graaff}, {Leja},
  {Marchesini}, {Miller}, {Nanayakkara}, {Oesch}, {Pan}, {Price}, {Setton},
  {Smit}, {Stefanon}, {Wang}, {Weaver}, {Whitaker}, {Williams}, \&
  {Zitrin}}]{Kokorev2023}
{Kokorev}, V., {Fujimoto}, S., {Labbe}, I., {et~al.} 2023, \apjl, 957, L7,
  \dodoi{10.3847/2041-8213/ad037a}

\bibitem[{{Kroupa}(2001)}]{Kroupa2001}
{Kroupa}, P. 2001, \mnras, 322, 231, \dodoi{10.1046/j.1365-8711.2001.04022.x}

\bibitem[{{Labb{\'e}} {et~al.}(2023){Labb{\'e}}, {van Dokkum}, {Nelson},
  {Bezanson}, {Suess}, {Leja}, {Brammer}, {Whitaker}, {Mathews}, {Stefanon}, \&
  {Wang}}]{Labbe2023}
{Labb{\'e}}, I., {van Dokkum}, P., {Nelson}, E., {et~al.} 2023, \nat, 616, 266,
  \dodoi{10.1038/s41586-023-05786-2}

\bibitem[{{Labbe} {et~al.}(2024){Labbe}, {Greene}, {Matthee}, {Treiber},
  {Kokorev}, {Miller}, {Kramarenko}, {Setton}, {Ma}, {Goulding}, {Bezanson},
  {Naidu}, {Williams}, {Atek}, {Brammer}, {Cutler}, {Chemerynska}, {Cloonan},
  {Dayal}, {de Graaff}, {Fudamoto}, {Fujimoto}, {Furtak}, {Glazebrook},
  {Heintz}, {Leja}, {Marchesini}, {Nanayakkara}, {Nelson}, {Oesch}, {Pan},
  {Price}, {Shivaei}, {Sobral}, {Suess}, {van Dokkum}, {Wang}, {Weaver},
  {Whitaker}, \& {Zitrin}}]{Labbe2024}
{Labbe}, I., {Greene}, J.~E., {Matthee}, J., {et~al.} 2024, arXiv e-prints,
  arXiv:2412.04557, \dodoi{10.48550/arXiv.2412.04557}

\bibitem[{{Labbe} {et~al.}(2025){Labbe}, {Greene}, {Bezanson}, {Fujimoto},
  {Furtak}, {Goulding}, {Matthee}, {Naidu}, {Oesch}, {Atek}, {Brammer},
  {Chemerynska}, {Coe}, {Cutler}, {Dayal}, {Feldmann}, {Franx}, {Glazebrook},
  {Leja}, {Maseda}, {Marchesini}, {Nanayakkara}, {Nelson}, {Pan}, {Papovich},
  {Price}, {Suess}, {Wang}, {Weaver}, {Whitaker}, {Williams}, \&
  {Zitrin}}]{Labbe2025}
{Labbe}, I., {Greene}, J.~E., {Bezanson}, R., {et~al.} 2025, \apj, 978, 92,
  \dodoi{10.3847/1538-4357/ad3551}

\bibitem[{{Larson} {et~al.}(2022){Larson}, {Finkelstein}, {Hutchison},
  {Papovich}, {Bagley}, {Dickinson}, {Rojas-Ruiz}, {Ferguson}, {Jung},
  {Giavalisco}, {Grazian}, {Pentericci}, \& {Tacchella}}]{Larson2022}
{Larson}, R.~L., {Finkelstein}, S.~L., {Hutchison}, T.~A., {et~al.} 2022, \apj,
  930, 104, \dodoi{10.3847/1538-4357/ac5dbd}

\bibitem[{{Larson} {et~al.}(2023){Larson}, {Finkelstein}, {Kocevski},
  {Hutchison}, {Trump}, {Arrabal Haro}, {Bromm}, {Cleri}, {Dickinson},
  {Fujimoto}, {Kartaltepe}, {Koekemoer}, {Papovich}, {Pirzkal}, {Tacchella},
  {Zavala}, {Bagley}, {Behroozi}, {Champagne}, {Cole}, {Jung}, {Morales},
  {Yang}, {Zhang}, {Zitrin}, {Amor{\'\i}n}, {Burgarella}, {Casey}, {Ch{\'a}vez
  Ortiz}, {Cox}, {Chworowsky}, {Fontana}, {Gawiser}, {Grazian}, {Grogin},
  {Harish}, {Hathi}, {Hirschmann}, {Holwerda}, {Juneau}, {Leung}, {Lucas},
  {McGrath}, {P{\'e}rez-Gonz{\'a}lez}, {Rigby}, {Seill{\'e}}, {Simons}, {de La
  Vega}, {Weiner}, {Wilkins}, {Yung}, \& {Ceers Team}}]{Larson2023}
{Larson}, R.~L., {Finkelstein}, S.~L., {Kocevski}, D.~D., {et~al.} 2023, \apjl,
  953, L29, \dodoi{10.3847/2041-8213/ace619}

\bibitem[{{Leonova} {et~al.}(2022){Leonova}, {Oesch}, {Qin}, {Naidu}, {Wyithe},
  {de Barros}, {Bouwens}, {Ellis}, {Endsley}, {Hutter}, {Illingworth},
  {Kerutt}, {Labb{\'e}}, {Laporte}, {Magee}, {Mutch}, {Roberts-Borsani},
  {Smit}, {Stark}, {Stefanon}, {Tacchella}, \& {Zitrin}}]{Leonova2022}
{Leonova}, E., {Oesch}, P.~A., {Qin}, Y., {et~al.} 2022, \mnras, 515, 5790,
  \dodoi{10.1093/mnras/stac1908}

\bibitem[{{Li} {et~al.}(2022){Li}, {Yang}, {Liu}, {Jing}, {He}, {Huang}, {Dai},
  {Sawicki}, {Arnouts}, {Gwyn}, {Moutard}, {Mo}, {Wang}, {Katsianis}, {Cui},
  {Han}, {Chiu}, {Gu}, \& {Xu}}]{Li2022}
{Li}, Q., {Yang}, X., {Liu}, C., {et~al.} 2022, \apj, 933, 9,
  \dodoi{10.3847/1538-4357/ac6e69}

\bibitem[{{Li} {et~al.}(2023){Li}, {Cai}, {Sun}, {Richard}, {Trebitsch},
  {Helton}, {Diego}, {Oguri}, {Foo}, {Lin}, {Bauer}, {Chen}, {Conselice},
  {Espada}, {Egami}, {Fan}, {Frye}, {Fudamoto}, {Perez-Gonzalez}, {Hainline},
  {Hsiao}, {Ji}, {Jin}, {Koekemoer}, {Kokorev}, {Kohno}, {Li}, {Lee}, {Magdis},
  {Willmer}, {Windhorst}, {Wu}, {Yan}, {Zhang}, {Zitrin}, {Zou}, {Bian},
  {Cheng}, {DeCoursey}, {Furtak}, {Steinhardt}, \& {Umehata}}]{Li2023}
{Li}, Z., {Cai}, Z., {Sun}, F., {et~al.} 2023, arXiv e-prints,
  arXiv:2310.09327, \dodoi{10.48550/arXiv.2310.09327}

\bibitem[{{Lim} {et~al.}(2024){Lim}, {Tacchella}, {Schaye}, {Schaller},
  {Helton}, {Kugel}, \& {Maiolino}}]{Lim2024}
{Lim}, S., {Tacchella}, S., {Schaye}, J., {et~al.} 2024, \mnras, 532, 4551,
  \dodoi{10.1093/mnras/stae1790}

\bibitem[{{Looser} {et~al.}(2024){Looser}, {D'Eugenio}, {Maiolino}, {Witstok},
  {Sandles}, {Curtis-Lake}, {Chevallard}, {Tacchella}, {Johnson}, {Baker},
  {Suess}, {Carniani}, {Ferruit}, {Arribas}, {Bonaventura}, {Bunker},
  {Cameron}, {Charlot}, {Curti}, {de Graaff}, {Maseda}, {Rawle}, {Rix}, {Del
  Pino}, {Smit}, {{\"U}bler}, {Willott}, {Alberts}, {Egami}, {Eisenstein},
  {Endsley}, {Hausen}, {Rieke}, {Robertson}, {Shivaei}, {Williams}, {Boyett},
  {Chen}, {Ji}, {Jones}, {Kumari}, {Nelson}, {Perna}, {Saxena}, \&
  {Scholtz}}]{Looser2024}
{Looser}, T.~J., {D'Eugenio}, F., {Maiolino}, R., {et~al.} 2024, \nat, 629, 53,
  \dodoi{10.1038/s41586-024-07227-0}

\bibitem[{{Ma} {et~al.}(2024){Ma}, {Sun}, {Cheng}, {Yan}, {Ling}, {Sun}, {Foo},
  {Egami}, {Diego}, {Cohen}, {Jansen}, {Summers}, {Windhorst}, {D'Silva},
  {Koekemoer}, {Coe}, {Conselice}, {Driver}, {Frye}, {Grogin}, {Marshall},
  {Nonino}, {Ortiz}, {Pirzkal}, {Robotham}, {Ryan}, {Willmer}, {Adams},
  {Hathi}, {Dole}, {Willner}, {Espada}, {Furtak}, {Hsiao}, {Li}, {Chen},
  {Jolly}, \& {Chen}}]{Ma2024}
{Ma}, Z., {Sun}, B., {Cheng}, C., {et~al.} 2024, \apj, 975, 87,
  \dodoi{10.3847/1538-4357/ad7b32}

\bibitem[{{Madau} \& {Dickinson}(2014)}]{Madau2014}
{Madau}, P., \& {Dickinson}, M. 2014, \araa, 52, 415,
  \dodoi{10.1146/annurev-astro-081811-125615}

\bibitem[{{Mason} {et~al.}(2023){Mason}, {Trenti}, \& {Treu}}]{Mason2023}
{Mason}, C.~A., {Trenti}, M., \& {Treu}, T. 2023, \mnras, 521, 497,
  \dodoi{10.1093/mnras/stad035}

\bibitem[{{Matthee} {et~al.}(2024){Matthee}, {Naidu}, {Brammer}, {Chisholm},
  {Eilers}, {Goulding}, {Greene}, {Kashino}, {Labbe}, {Lilly}, {Mackenzie},
  {Oesch}, {Weibel}, {Wuyts}, {Xiao}, {Bordoloi}, {Bouwens}, {van Dokkum},
  {Illingworth}, {Kramarenko}, {Maseda}, {Mason}, {Meyer}, {Nelson}, {Reddy},
  {Shivaei}, {Simcoe}, \& {Yue}}]{Matthee2024}
{Matthee}, J., {Naidu}, R.~P., {Brammer}, G., {et~al.} 2024, \apj, 963, 129,
  \dodoi{10.3847/1538-4357/ad2345}

\bibitem[{{McQuinn}(2016)}]{McQuinn2016}
{McQuinn}, M. 2016, \araa, 54, 313, \dodoi{10.1146/annurev-astro-082214-122355}

\bibitem[{{Meyer} {et~al.}(2024){Meyer}, {Oesch}, {Giovinazzo}, {Weibel},
  {Brammer}, {Matthee}, {Naidu}, {Bouwens}, {Chisholm}, {Covelo-Paz},
  {Fudamoto}, {Maseda}, {Nelson}, {Shivaei}, {Xiao}, {Herard-Demanche},
  {Illingworth}, {Kerutt}, {Kramarenko}, {Labbe}, {Leonova}, {Magee},
  {Matharu}, {Prieto Lyon}, {Reddy}, {Schaerer}, {Shapley}, {Stefanon},
  {Wozniak}, \& {Wuyts}}]{Meyer2024}
{Meyer}, R.~A., {Oesch}, P.~A., {Giovinazzo}, E., {et~al.} 2024, \mnras, 535,
  1067, \dodoi{10.1093/mnras/stae2353}

\bibitem[{{Morishita} {et~al.}(2023){Morishita}, {Roberts-Borsani}, {Treu},
  {Brammer}, {Mason}, {Trenti}, {Vulcani}, {Wang}, {Acebron}, {Bah{\'e}},
  {Bergamini}, {Boyett}, {Bradac}, {Calabr{\`o}}, {Castellano}, {Chen}, {De
  Lucia}, {Filippenko}, {Fontana}, {Glazebrook}, {Grillo}, {Henry}, {Jones},
  {Kelly}, {Koekemoer}, {Leethochawalit}, {Lu}, {Marchesini}, {Mascia},
  {Mercurio}, {Merlin}, {Metha}, {Nanayakkara}, {Nonino}, {Paris},
  {Pentericci}, {Rosati}, {Santini}, {Strait}, {Vanzella}, {Windhorst}, \&
  {Xie}}]{Morishita2023}
{Morishita}, T., {Roberts-Borsani}, G., {Treu}, T., {et~al.} 2023, \apjl, 947,
  L24, \dodoi{10.3847/2041-8213/acb99e}

\bibitem[{{Morishita} {et~al.}(2024){Morishita}, {Liu}, {Stiavelli}, {Treu},
  {Trenti}, {Chartab}, {Roberts-Borsani}, {Vulcani}, {Bergamini}, {Castellano},
  \& {Grillo}}]{Morishita2024}
{Morishita}, T., {Liu}, Z., {Stiavelli}, M., {et~al.} 2024, arXiv e-prints,
  arXiv:2408.10980, \dodoi{10.48550/arXiv.2408.10980}

\bibitem[{{Morishita} {et~al.}(2025){Morishita}, {Stiavelli}, {Vanzella},
  {Bergamini}, {Boyett}, {Chiaberge}, {Grillo}, {Leethochawalit}, {Messa},
  {Roberts-Borsani}, {Rosati}, \& {Shajib}}]{Morishita2025}
{Morishita}, T., {Stiavelli}, M., {Vanzella}, E., {et~al.} 2025, arXiv
  e-prints, arXiv:2501.11879, \dodoi{10.48550/arXiv.2501.11879}

\bibitem[{{Naidu} {et~al.}(2022){Naidu}, {Oesch}, {van Dokkum}, {Nelson},
  {Suess}, {Brammer}, {Whitaker}, {Illingworth}, {Bouwens}, {Tacchella},
  {Matthee}, {Allen}, {Bezanson}, {Conroy}, {Labbe}, {Leja}, {Leonova},
  {Magee}, {Price}, {Setton}, {Strait}, {Stefanon}, {Toft}, {Weaver}, \&
  {Weibel}}]{Naidu2022}
{Naidu}, R.~P., {Oesch}, P.~A., {van Dokkum}, P., {et~al.} 2022, \apjl, 940,
  L14, \dodoi{10.3847/2041-8213/ac9b22}

\bibitem[{{Nakajima} {et~al.}(2023){Nakajima}, {Ouchi}, {Isobe}, {Harikane},
  {Zhang}, {Ono}, {Umeda}, \& {Oguri}}]{Nakajima2023}
{Nakajima}, K., {Ouchi}, M., {Isobe}, Y., {et~al.} 2023, \apjs, 269, 33,
  \dodoi{10.3847/1538-4365/acd556}

\bibitem[{{Oesch} {et~al.}(2023){Oesch}, {Brammer}, {Naidu}, {Bouwens},
  {Chisholm}, {Illingworth}, {Matthee}, {Nelson}, {Qin}, {Reddy}, {Shapley},
  {Shivaei}, {van Dokkum}, {Weibel}, {Whitaker}, {Wuyts}, {Covelo-Paz},
  {Endsley}, {Fudamoto}, {Giovinazzo}, {Herard-Demanche}, {Kerutt},
  {Kramarenko}, {Labbe}, {Leonova}, {Lin}, {Magee}, {Marchesini}, {Maseda},
  {Mason}, {Matharu}, {Meyer}, {Neufeld}, {Prieto Lyon}, {Schaerer}, {Sharma},
  {Shuntov}, {Smit}, {Stefanon}, {Wyithe}, \& {Xiao}}]{Oesch2023}
{Oesch}, P.~A., {Brammer}, G., {Naidu}, R.~P., {et~al.} 2023, \mnras, 525,
  2864, \dodoi{10.1093/mnras/stad2411}

\bibitem[{{Oke} \& {Gunn}(1983)}]{1983ApJ...266..713O}
{Oke}, J.~B., \& {Gunn}, J.~E. 1983, \apj, 266, 713, \dodoi{10.1086/160817}

\bibitem[{{Planck Collaboration} {et~al.}(2016){Planck Collaboration}, {Adam},
  {Aghanim}, {Ashdown}, {Aumont}, {Baccigalupi}, {Ballardini}, {Banday},
  {Barreiro}, {Bartolo}, {Basak}, {Battye}, {Benabed}, {Bernard}, {Bersanelli},
  {Bielewicz}, {Bock}, {Bonaldi}, {Bonavera}, {Bond}, {Borrill}, {Bouchet},
  {Boulanger}, {Bucher}, {Burigana}, {Calabrese}, {Cardoso}, {Carron},
  {Chiang}, {Colombo}, {Combet}, {Comis}, {Couchot}, {Coulais}, {Crill},
  {Curto}, {Cuttaia}, {Davis}, {de Bernardis}, {de Rosa}, {de Zotti},
  {Delabrouille}, {Di Valentino}, {Dickinson}, {Diego}, {Dor{\'e}}, {Douspis},
  {Ducout}, {Dupac}, {Elsner}, {En{\ss}lin}, {Eriksen}, {Falgarone}, {Fantaye},
  {Finelli}, {Forastieri}, {Frailis}, {Fraisse}, {Franceschi}, {Frolov},
  {Galeotta}, {Galli}, {Ganga}, {G{\'e}nova-Santos}, {Gerbino}, {Ghosh},
  {Gonz{\'a}lez-Nuevo}, {G{\'o}rski}, {Gruppuso}, {Gudmundsson}, {Hansen},
  {Helou}, {Henrot-Versill{\'e}}, {Herranz}, {Hivon}, {Huang}, {Ili{\'c}},
  {Jaffe}, {Jones}, {Keih{\"a}nen}, {Keskitalo}, {Kisner}, {Knox},
  {Krachmalnicoff}, {Kunz}, {Kurki-Suonio}, {Lagache}, {L{\"a}hteenm{\"a}ki},
  {Lamarre}, {Langer}, {Lasenby}, {Lattanzi}, {Lawrence}, {Le Jeune},
  {Levrier}, {Lewis}, {Liguori}, {Lilje}, {L{\'o}pez-Caniego}, {Ma},
  {Mac{\'\i}as-P{\'e}rez}, {Maggio}, {Mangilli}, {Maris}, {Martin},
  {Mart{\'\i}nez-Gonz{\'a}lez}, {Matarrese}, {Mauri}, {McEwen}, {Meinhold},
  {Melchiorri}, {Mennella}, {Migliaccio}, {Miville-Desch{\^e}nes}, {Molinari},
  {Moneti}, {Montier}, {Morgante}, {Moss}, {Naselsky}, {Natoli}, {Oxborrow},
  {Pagano}, {Paoletti}, {Partridge}, {Patanchon}, {Patrizii}, {Perdereau},
  {Perotto}, {Pettorino}, {Piacentini}, {Plaszczynski}, {Polastri}, {Polenta},
  {Puget}, {Rachen}, {Racine}, {Reinecke}, {Remazeilles}, {Renzi}, {Rocha},
  {Rossetti}, {Roudier}, {Rubi{\~n}o-Mart{\'\i}n}, {Ruiz-Granados}, {Salvati},
  {Sandri}, {Savelainen}, {Scott}, {Sirri}, {Sunyaev}, {Suur-Uski}, {Tauber},
  {Tenti}, {Toffolatti}, {Tomasi}, {Tristram}, {Trombetti}, {Valiviita}, {Van
  Tent}, {Vielva}, {Villa}, {Vittorio}, {Wandelt}, {Wehus}, {White}, {Zacchei},
  \& {Zonca}}]{Planck2016}
{Planck Collaboration}, {Adam}, R., {Aghanim}, N., {et~al.} 2016, \aap, 596,
  A108, \dodoi{10.1051/0004-6361/201628897}

\bibitem[{{Richard} {et~al.}(2021){Richard}, {Claeyssens}, {Lagattuta},
  {Guaita}, {Bauer}, {Pello}, {Carton}, {Bacon}, {Soucail}, {Lyon}, {Kneib},
  {Mahler}, {Cl{\'e}ment}, {Mercier}, {Variu}, {Tamone}, {Ebeling}, {Schmidt},
  {Nanayakkara}, {Maseda}, {Weilbacher}, {Bouch{\'e}}, {Bouwens}, {Wisotzki},
  {de la Vieuville}, {Martinez}, \& {Patr{\'\i}cio}}]{richard_2021}
{Richard}, J., {Claeyssens}, A., {Lagattuta}, D., {et~al.} 2021, \aap, 646,
  A83, \dodoi{10.1051/0004-6361/202039462}

\bibitem[{{Roberts-Borsani} {et~al.}(2024){Roberts-Borsani}, {Treu}, {Shapley},
  {Fontana}, {Pentericci}, {Castellano}, {Morishita}, {Bergamini}, \&
  {Rosati}}]{Roberts-Borsani2024}
{Roberts-Borsani}, G., {Treu}, T., {Shapley}, A., {et~al.} 2024, \apj, 976,
  193, \dodoi{10.3847/1538-4357/ad85d3}

\bibitem[{{Robertson}(2022)}]{Robertson2022}
{Robertson}, B.~E. 2022, \araa, 60, 121,
  \dodoi{10.1146/annurev-astro-120221-044656}

\bibitem[{{Rodr{\'\i}guez-Puebla} {et~al.}(2016){Rodr{\'\i}guez-Puebla},
  {Behroozi}, {Primack}, {Klypin}, {Lee}, \& {Hellinger}}]{RP16}
{Rodr{\'\i}guez-Puebla}, A., {Behroozi}, P., {Primack}, J., {et~al.} 2016,
  \mnras, 462, 893, \dodoi{10.1093/mnras/stw1705}

\bibitem[{{Saxena} {et~al.}(2023){Saxena}, {Robertson}, {Bunker}, {Endsley},
  {Cameron}, {Charlot}, {Simmonds}, {Tacchella}, {Witstok}, {Willott},
  {Carniani}, {Curtis-Lake}, {Ferruit}, {Jakobsen}, {Arribas}, {Chevallard},
  {Curti}, {D'Eugenio}, {De Graaff}, {Jones}, {Looser}, {Maseda}, {Rawle},
  {Rix}, {Del Pino}, {Smit}, {{\"U}bler}, {Eisenstein}, {Hainline}, {Hausen},
  {Johnson}, {Rieke}, {Williams}, {Willmer}, {Baker}, {Bhatawdekar}, {Bowler},
  {Boyett}, {Chen}, {Egami}, {Ji}, {Kumari}, {Nelson}, {Perna}, {Sandles},
  {Scholtz}, \& {Shivaei}}]{Saxena2023}
{Saxena}, A., {Robertson}, B.~E., {Bunker}, A.~J., {et~al.} 2023, \aap, 678,
  A68, \dodoi{10.1051/0004-6361/202346245}

\bibitem[{{Schouws} {et~al.}(2024){Schouws}, {Bouwens}, {Ormerod}, {Smit},
  {Algera}, {Sommovigo}, {Hodge}, {Ferrara}, {Oesch}, {Rowland}, {van Leeuwen},
  {Stefanon}, {Herard-Demanche}, {Fudamoto}, {R{\"o}ttgering}, \& {van der
  Werf}}]{Schouws2024}
{Schouws}, S., {Bouwens}, R.~J., {Ormerod}, K., {et~al.} 2024, arXiv e-prints,
  arXiv:2409.20549, \dodoi{10.48550/arXiv.2409.20549}

\bibitem[{{Somerville} \& {Dav{\'e}}(2015)}]{Somerville2015}
{Somerville}, R.~S., \& {Dav{\'e}}, R. 2015, \araa, 53, 51,
  \dodoi{10.1146/annurev-astro-082812-140951}

\bibitem[{{Sun} {et~al.}(2025){Sun}, {Fudamoto}, {Lin}, {Helton}, {Hsiao},
  {Egami}, \& et~al.}]{Sun2025}
{Sun}, F., {Fudamoto}, Y., {Lin}, X., {et~al.} 2025

\bibitem[{{Sun} {et~al.}(2023){Sun}, {Egami}, {Pirzkal}, {Rieke}, {Baum},
  {Boyer}, {Boyett}, {Bunker}, {Cameron}, {Curti}, {Eisenstein}, {Gennaro},
  {Greene}, {Jaffe}, {Kelly}, {Koekemoer}, {Kumari}, {Maiolino}, {Maseda},
  {Perna}, {Rest}, {Robertson}, {Schlawin}, {Smit}, {Stansberry}, {Sunnquist},
  {Tacchella}, {Williams}, \& {Willmer}}]{Sun2023}
{Sun}, F., {Egami}, E., {Pirzkal}, N., {et~al.} 2023, \apj, 953, 53,
  \dodoi{10.3847/1538-4357/acd53c}

\bibitem[{{Sun} {et~al.}(2024){Sun}, {Helton}, {Egami}, {Hainline}, {Rieke},
  {Willmer}, {Eisenstein}, {Johnson}, {Rieke}, {Robertson}, {Tacchella},
  {Alberts}, {Baker}, {Bhatawdekar}, {Boyett}, {Bunker}, {Charlot}, {Chen},
  {Chevallard}, {Curtis-Lake}, {Danhaive}, {DeCoursey}, {Ji}, {Lyu},
  {Maiolino}, {Rujopakarn}, {Sandles}, {Shivaei}, {{\"U}bler}, {Willott}, \&
  {Witstok}}]{Sun2024}
{Sun}, F., {Helton}, J.~M., {Egami}, E., {et~al.} 2024, \apj, 961, 69,
  \dodoi{10.3847/1538-4357/ad07e3}

\bibitem[{{Tamura} {et~al.}(2019){Tamura}, {Mawatari}, {Hashimoto}, {Inoue},
  {Zackrisson}, {Christensen}, {Binggeli}, {Matsuda}, {Matsuo}, {Takeuchi},
  {Asano}, {Sunaga}, {Shimizu}, {Okamoto}, {Yoshida}, {Lee}, {Shibuya},
  {Taniguchi}, {Umehata}, {Hatsukade}, {Kohno}, \& {Ota}}]{Tamura2019}
{Tamura}, Y., {Mawatari}, K., {Hashimoto}, T., {et~al.} 2019, \apj, 874, 27,
  \dodoi{10.3847/1538-4357/ab0374}

\bibitem[{{Tamura} {et~al.}(2023){Tamura}, {C. Bakx}, {Inoue}, {Hashimoto},
  {Tokuoka}, {Imamura}, {Hatsukade}, {Lee}, {Moriwaki}, {Okamoto}, {Ota},
  {Umehata}, {Yoshida}, {Zackrisson}, {Hagimoto}, {Matsuo}, {Shimizu},
  {Sugahara}, \& {Takeuchi}}]{Tamura2023}
{Tamura}, Y., {C. Bakx}, T. J.~L., {Inoue}, A.~K., {et~al.} 2023, \apj, 952, 9,
  \dodoi{10.3847/1538-4357/acd637}

\bibitem[{{Tilvi} {et~al.}(2020){Tilvi}, {Malhotra}, {Rhoads}, {Coughlin},
  {Zheng}, {Finkelstein}, {Veilleux}, {Mobasher}, {Wang}, {Probst}, {Swaters},
  {Hibon}, {Joshi}, {Zabl}, {Jiang}, {Pharo}, \& {Yang}}]{Tilvi2020}
{Tilvi}, V., {Malhotra}, S., {Rhoads}, J.~E., {et~al.} 2020, \apjl, 891, L10,
  \dodoi{10.3847/2041-8213/ab75ec}

\bibitem[{{Toshikawa} {et~al.}(2014){Toshikawa}, {Kashikawa}, {Overzier},
  {Shibuya}, {Ishikawa}, {Ota}, {Shimasaku}, {Tanaka}, {Hayashi}, {Niino}, \&
  {Onoue}}]{Toshikawa2014}
{Toshikawa}, J., {Kashikawa}, N., {Overzier}, R., {et~al.} 2014, \apj, 792, 15,
  \dodoi{10.1088/0004-637X/792/1/15}

\bibitem[{{Tremonti} {et~al.}(2004){Tremonti}, {Heckman}, {Kauffmann},
  {Brinchmann}, {Charlot}, {White}, {Seibert}, {Peng}, {Schlegel}, {Uomoto},
  {Fukugita}, \& {Brinkmann}}]{Tremonti2004}
{Tremonti}, C.~A., {Heckman}, T.~M., {Kauffmann}, G., {et~al.} 2004, \apj, 613,
  898, \dodoi{10.1086/423264}

\bibitem[{{Villa-V{\'e}lez} {et~al.}(2021){Villa-V{\'e}lez}, {Buat},
  {Theul{\'e}}, {Boquien}, \& {Burgarella}}]{Villa-Velez2021}
{Villa-V{\'e}lez}, J.~A., {Buat}, V., {Theul{\'e}}, P., {Boquien}, M., \&
  {Burgarella}, D. 2021, \aap, 654, A153, \dodoi{10.1051/0004-6361/202140890}

\bibitem[{{Wang} {et~al.}(2024){Wang}, {Leja}, {Labb{\'e}}, {Bezanson},
  {Whitaker}, {Brammer}, {Furtak}, {Weaver}, {Price}, {Zitrin}, {Atek}, {Coe},
  {Cutler}, {Dayal}, {van Dokkum}, {Feldmann}, {Marchesini}, {Franx},
  {F{\"o}rster Schreiber}, {Fujimoto}, {Geha}, {Glazebrook}, {de Graaff},
  {Greene}, {Juneau}, {Kassin}, {Kriek}, {Khullar}, {Maseda}, {Mowla},
  {Muzzin}, {Nanayakkara}, {Nelson}, {Oesch}, {Pacifici}, {Pan}, {Papovich},
  {Setton}, {Shapley}, {Smit}, {Stefanon}, {Suess}, {Taylor}, \&
  {Williams}}]{Wang2024}
{Wang}, B., {Leja}, J., {Labb{\'e}}, I., {et~al.} 2024, \apjs, 270, 12,
  \dodoi{10.3847/1538-4365/ad0846}

\bibitem[{{Wechsler} \& {Tinker}(2018)}]{Wechsler2018}
{Wechsler}, R.~H., \& {Tinker}, J.~L. 2018, \araa, 56, 435,
  \dodoi{10.1146/annurev-astro-081817-051756}

\bibitem[{{Whitler} {et~al.}(2024){Whitler}, {Stark}, {Endsley}, {Chen},
  {Mason}, {Topping}, \& {Charlot}}]{Whitler2024}
{Whitler}, L., {Stark}, D.~P., {Endsley}, R., {et~al.} 2024, \mnras, 529, 855,
  \dodoi{10.1093/mnras/stae516}

\bibitem[{{Wise}(2019)}]{Wise2019}
{Wise}, J.~H. 2019, Contemporary Physics, 60, 145,
  \dodoi{10.1080/00107514.2019.1631548}

\bibitem[{{Witstok} {et~al.}(2024{\natexlab{a}}){Witstok}, {Smit}, {Saxena},
  {Jones}, {Helton}, {Sun}, {Maiolino}, {Kumari}, {Stark}, {Bunker}, {Arribas},
  {Baker}, {Bhatawdekar}, {Boyett}, {Cameron}, {Carniani}, {Charlot},
  {Chevallard}, {Curti}, {Curtis-Lake}, {Eisenstein}, {Endsley}, {Hainline},
  {Ji}, {Johnson}, {Looser}, {Nelson}, {Perna}, {Rix}, {Robertson}, {Sandles},
  {Scholtz}, {Simmonds}, {Tacchella}, {{\"U}bler}, {Williams}, {Willmer}, \&
  {Willott}}]{Witstok2024a}
{Witstok}, J., {Smit}, R., {Saxena}, A., {et~al.} 2024{\natexlab{a}}, \aap,
  682, A40, \dodoi{10.1051/0004-6361/202347176}

\bibitem[{{Witstok} {et~al.}(2024{\natexlab{b}}){Witstok}, {Maiolino}, {Smit},
  {Jones}, {Bunker}, {Helton}, {Johnson}, {Tacchella}, {Saxena}, {Arribas},
  {Bhatawdekar}, {Boyett}, {Cameron}, {Cargile}, {Carniani}, {Charlot},
  {Chevallard}, {Curti}, {Curtis-Lake}, {D'Eugenio}, {Eisenstein}, {Hainline},
  {Hausen}, {Kumari}, {Laseter}, {Maseda}, {Rieke}, {Robertson}, {Scholtz},
  {Shivaei}, {Williams}, {Willmer}, \& {Willott}}]{Witstok2024b}
{Witstok}, J., {Maiolino}, R., {Smit}, R., {et~al.} 2024{\natexlab{b}}, \mnras,
  \dodoi{10.1093/mnras/stae2535}

\bibitem[{{Witstok} {et~al.}(2025){Witstok}, {Maiolino}, {Smit}, {Jones},
  {Bunker}, {Helton}, {Johnson}, {Tacchella}, {Saxena}, {Arribas},
  {Bhatawdekar}, {Boyett}, {Cameron}, {Cargile}, {Carniani}, {Charlot},
  {Chevallard}, {Curti}, {Curtis-Lake}, {D'Eugenio}, {Eisenstein}, {Hainline},
  {Hausen}, {Kumari}, {Laseter}, {Maseda}, {Rieke}, {Robertson}, {Scholtz},
  {Shivaei}, {Williams}, {Willmer}, \& {Willott}}]{Witstok2025}
---. 2025, \mnras, 536, 27, \dodoi{10.1093/mnras/stae2535}

\bibitem[{{Witten} {et~al.}(2025){Witten}, {McClymont}, {Laporte},
  {Roberts-Borsani}, {Sijacki}, {Tacchella}, {Simmonds}, {Katz}, {Ellis},
  {Witstok}, {Maiolino}, {Ji}, {Hayes}, {Looser}, \& {D'Eugenio}}]{Witten2025}
{Witten}, C., {McClymont}, W., {Laporte}, N., {et~al.} 2025, \mnras, 537, 112,
  \dodoi{10.1093/mnras/staf001}

\bibitem[{{Xiao} {et~al.}(2024){Xiao}, {Oesch}, {Elbaz}, {Bing}, {Nelson},
  {Weibel}, {Illingworth}, {van Dokkum}, {Naidu}, {Daddi}, {Bouwens},
  {Matthee}, {Wuyts}, {Chisholm}, {Brammer}, {Dickinson}, {Magnelli}, {Leroy},
  {Schaerer}, {Herard-Demanche}, {Lim}, {Barrufet}, {Endsley}, {Fudamoto},
  {G{\'o}mez-Guijarro}, {Gottumukkala}, {Labb{\'e}}, {Magee}, {Marchesini},
  {Maseda}, {Qin}, {Reddy}, {Shapley}, {Shivaei}, {Shuntov}, {Stefanon},
  {Whitaker}, \& {Wyithe}}]{Xiao2024}
{Xiao}, M., {Oesch}, P.~A., {Elbaz}, D., {et~al.} 2024, \nat, 635, 311,
  \dodoi{10.1038/s41586-024-08094-5}

\end{thebibliography}
\bibliographystyle{aasjournal}



\end{document}